\documentclass[aps,prd,nofootinbib,showpacs,floatfix]{revtex4}

\usepackage{graphicx}

\begin{document}
%
%
\def\beqa{\begin{eqnarray}}
\def\eeqa{\end{eqnarray}}
\def\beq{\begin{equation}}
\def\eeq{\end{equation}}
\def\p{\partial}
\def\vp{\varphi}
\def\d{\delta}
\def\ulap{\underline\Delta}
\def\unab{\overline\nabla}
\def\wg#1{\mbox{\boldmath ${#1}$}}
\def\w#1{\bf #1}

\title{Various features of quasiequilibrium sequences of
  binary neutron stars in general relativity}

\author{Keisuke Taniguchi}
\email[]{keisuke@provence.c.u-tokyo.ac.jp}
\affiliation{Department of Earth Science and Astronomy,
  Graduate School of Arts and Sciences, \\
  University of Tokyo, Komaba, Meguro, Tokyo 153-8902, Japan
}

\author{Eric Gourgoulhon}
\email[]{Eric.Gourgoulhon@obspm.fr}
\affiliation{Laboratoire de l'Univers et de ses Th\'eories,
UMR 8102 du C.N.R.S., Observatoire de Paris, F-92195 Meudon Cedex, France}

\date{13 October 2003}

\begin{abstract}

Quasiequilibrium sequences of binary neutron stars are
numerically calculated
in the framework of the Isenberg-Wilson-Mathews (IWM) approximation
of general relativity.
The results are presented for both rotation states of synchronized spins
and irrotational motion, the latter being considered as
the realistic one for binary neutron stars just prior to the merger.
We assume a polytropic equation of state 
and compute several evolutionary sequences of binary systems
composed of different-mass stars as well as identical-mass stars
with adiabatic indices $\gamma=2.5, 2.25, 2$, and $1.8$.
From our results, we propose as a conjecture that
if the turning point of binding energy (and total angular momentum)
locating the innermost stable circular orbit (ISCO)
is found in Newtonian gravity for some value of the adiabatic index
$\gamma_0$, that of the ADM mass (and total angular momentum)
should exist in the IWM approximation of general relativity
for the same value of the adiabatic index.

\end{abstract}



\maketitle

\section{Introduction} \label{s:intro}

The final stage of coalescing binary neutron stars is 
one of the most promising sources of gravitational waves
for ground-based laser interferometers
such as GEO600, LIGO, TAMA300, and VIRGO \cite{CutleT02}.
It is also considered as one of the candidates for short-duration
gamma-ray burst sources \cite{NarayPP92}.
With accurate templates of gravitational waves from
coalescing binary neutron stars,
it may be possible to extract informations from signals
observed by the above interferometers \cite{Tagos01},
through the matched filtering technique \cite{TanakT00}.
Therefore, it is essential to theoretically predict the wave form
of gravitational radiation
as well as to understand the details of the physics of binary neutron stars.
With such a motivation, many efforts have been invested in this topic
(some recent reviews can be found in \cite{BaumgS03,Blanc02b})
When summarizing these researches,
it is convenient to separate the evolution of binary neutron stars 
into three phases.
The first one is the {\em pointlike inspiral} in which the separation
between the two stars is much larger than the radius of one star,
and the binary system evolves adiabatically.
In this stage, the post-Newtonian expansion works excellently,
and accurate gravitational wave forms
have been calculated \cite{BlancFIJ02,Blanc02b,DamouIJS03}.
The second phase is the {\em intermediate} one or {\em hydrodynamical
inspiral} in which 
the separation between the two components is as small as a few times of
the radius of one star.
This phase is also inspiraling,
the time scale of the orbital shrink being still larger than
the orbital period, hence the qualifier of {\em quasiequilibrium}
given to it. 
What differs this phase from the pointlike inspiral is that
the tidal deformation of the stars is no longer negligible,
making necessary a hydrodynamical treatment, in addition to 
general relativity.
The final phase is the {\em merger}, in which the two stars
coalesces dynamically.
The first successful merger computations have been obtained 
by Shibata et al. four years ago \cite{Shiba99,ShibaU00,ShibaTU03},
and various groups are currently working on this subject
 \cite{OoharN99,KawamON03,FontGIMRSSST02,DuezMSB03}.

In the present paper, we focus on the intermediate phase (hydrodynamical
inspiral).
This phase is interesting because one might get informations
on the equation of state of nuclear matter \cite{FaberGRT02}.
Furthermore, the most realistic initial data for computing the merger 
are given at the end point
of sequences in the hydrodynamical inspiral \cite{ShibaTU03}
\footnote{Miller and Suen pointed out that the circular orbit condition
as an initial data is not a good approximation for {\em corotating}
neutron star binary systems \cite{MilleS03,Miller03}. However,
as we describe below, the realistic rotation state of
binary neutron stars prior to merger is irrotational flow,
and such initial data have been used in Refs.~\cite{ShibaU00,ShibaTU03}.
Furthermore, in Ref.~\cite{ShibaTU03},
initial data are given at some separations slightly larger than
the innermost stable circular orbit.
We hope that they do not deviate too much from the realistic inspiral
for irrotational binary neutron stars.}.

Various approaches to compute binary neutron star models in the 
hydrodynamical inspiral phase have been considered:
{\em (i)} (semi-)analytical \cite{LaiRS93,TanigN00} and
{\em (ii)} numerical \cite{HachiE84a,HachiE84b,UryuE98,TanigGB01,TanigG02a}
in  Newtonian gravity,
{\em (iii)} (semi-)analytical \cite{LombaRS97,TanigS97,ShibaT97,Tanig99} and
{\em (iv)} numerical \cite{Shiba96} in the post-Newtonian approximation,
and {\em (v)} numerical in the framework of general relativity
\cite{BaumgCSST97,MarroMW98,BonazGM99a,GourGTMB01,TanigG02b,UryuE00,UryuSE00,UsuiUE99,UsuiE02,MarroS03}.
Binary neutron stars gradually decrease their orbital
separations as a result of the emission of gravitational radiation.
However, it is still impossible to integrate the Einstein's equations
for the thousands of orbits of the hydrodynamical inspiral.
Fortunately the time scale of the orbit shrink being much longer than
the orbital period, the hydrodynamical inspiral phase can be approximated by
a sequence of steady-state (quasiequilibrium) configurations.
In order to combine general relativity and quasiequilibrium,
we adopt the conformal-flatness condition for the spatial part of the metric
--- the so-called Isenberg-Wilson-Mathews (IWM) approximation
of general relativity \cite{Isenb78,IsenbN80,WilsoM89} (see
\cite{FriedUS02} for a discussion).
This treatment postpones one of the goals of the study of the 
hydrodynamical inspiral, i.e. the theoretical prediction of the wave form of 
gravitational radiation\footnote{see however Refs.~\cite{DuezBS01,ShibaU01}
for the computation of gravitational waves from IWM configurations.}.
Accordingly, we focus on the evolution of various physical parameters
during the hydrodynamical inspiral and investigate the nature 
of the innermost orbit (mass-shedding or orbital instability). 

To get a  quasiequilibrium configuration of binary neutron stars,
we have to specify the rotation state of the system.
About ten years ago, Kochaneck \cite{Kocha92}
and Bildsten and Cutler \cite{BildsC92} 
concluded that the irrotational flow is much more realistic
than the synchronized rotation, 
because the shear viscosity of nuclear matter is far too low to synchronize
the spins of the stars with the orbital motion by the dynamical merging.
The formulation for solving quasiequilibrium configurations of irrotational
binary neutron stars in general relativity
has been proposed by several authors
\cite{BonazGM97,Asada98,Shiba98,Teuko98}.
In the present work, we have calculated not only 
irrotational configurations but also synchronized ones
in order to exhibit the differences between these two extreme states.
Beside the type of rotation, some equation of state for neutron 
star matter must be specified to get a binary model.
Although some realistic equation of state arising from nuclear physics
(see e.g. \cite{Haens03}) has to be used,
we adopt here the polytropic one for simplicity. 
Nevertheless we vary the adiabatic index in the range $\gamma=1.8$
to $3$  to cover the large range of nuclear equations of state
published in the literature. 
Quasiequilibrium sequences of binary neutron stars
based on a selection of 
realistic equations of state will be presented in a future work.

Under the above assumptions, 
we have computed quasiequilibrium sequences of binary systems composed
of neutron stars with different masses. In the three known 
binary neutron star systems (observed as binary pulsars with a large
mass companion), the two components have almost the same mass:
the relative difference between the masses of the two stars are 
4 \% for PSR B1913+16 \cite{ThorsC99,TayloW89}, less than 0.1 \%
for PSR B1534+12 \cite{ThorsC99,Stairs98}, and about 1 \%
for PSR B2127+11C \cite{ThorsC99}.
Until now, all studies on quasiequilibrium sequences of binary neutron stars
in the framework of general relativity
have dealt with only identical-mass binary systems,
except for our previous work \cite{TanigG02b} (hereafter Paper III).
In that article we have also presented quasiequilibrium sequences
of binary neutron stars with a polytropic equation of state.
However, we limited ourselves to the adiabatic index $\gamma=2$,
because it was the first step to relativistic 
different-mass binary systems.
In the present article, we enlarge the range of adiabatic index,
investigating softer equations of state ($\gamma=1.8$) and stiffer ones
($\gamma=2.25$, $2.5$ and $3$).

The plan of the article is as follows:
Section~\ref{s:method} provides a brief summary of the assumptions, 
basic equations,
and numerical methods which we adopt.
Some code tests not shown in previous articles of this series 
are given in Sec.~\ref{s:tests},
and the numerical results are presented in Sec.~\ref{s:results}.
Section~\ref{s:discussion} is devoted to the discussions and
we conclude the paper with a summary in Sec.~\ref{s:summary}.
Through the paper, we use units in which $c=G=1$,
where $c$ is the speed of light and $G$ the gravitational constant.

\section{Method} \label{s:method}

The details about our method and basic equations
have been already given in previous papers
of this series
(Refs. \cite{GourGTMB01} and \cite{TanigGB01}, hereafter Paper I and II 
respectively). 
We refer the interested reader to those articles
and simply summarize here the basic assumptions, the basic equations and
our numerical method.

\subsection{Assumptions}

Firstly, we assume the binary system to be
in a {\em quasiequilibrium} state.
This assumption comes from the fact that the time scale $t_{\rm GW}$
of the orbital shrink driven by the emission of gravitational radiation
is longer than the orbital period $P_{\rm orb}$. Indeed the following
relation holds:
\beq
  {t_{\rm GW} \over P_{\rm orb}} \simeq 1.1 \Bigl( {d \over 6M_{\rm tot}}
  \Bigr)^{5/2} \Bigl( {M_{\rm tot} \over 4\mu} \Bigr),
\eeq
where $d$, $M_{\rm tot}$, and $\mu$ denote
respectively the separation between two stars,
the total mass, and the reduced mass. Hence for an equal-mass system,
$t_{\rm GW} > P_{\rm orb}$ for $d >6M_{\rm tot}$.
It is well known that apart from shrinking the orbits, the reaction
to gravitational radiation circularizes them. This implies a continuous
spacetime symmetry, called {\em helical symmetry} \cite{BonazGM97,FriedUS02}
and represented by the Killing vector:
\beq
  \wg{l} = {\p \over \p t} +\Omega {\p \over \p \vp},
\eeq
where $\Omega$ is the orbital angular velocity and 
$\p/\p t$ and $\p/\p \vp$ are the natural frame vectors 
associated with the time
coordinate $t$ and the azimuthal coordinate $\vp$ of an asymptotic
inertial observer.

The second assumption is applied to the stress-energy tensor,
which we require to have the {\em perfect fluid} form:
\beq
  T_{\mu \nu} = (e+p) u_{\mu} u_{\nu} + p \, g_{\mu \nu},
\eeq
where $e$ denotes the fluid proper energy density,
$p$ the fluid pressure, $u_{\mu}$ the fluid 4-velocity,
and $g_{\mu \nu}$ the spacetime metric. This constitutes a very good
approximation for neutron star matter. 

The third assumption concerns the fluid motion inside each star.
We consider both cases of {\em synchronized} (also called {\em corotating})
motion and {\em irrotational} flow. Note that while only the latter
state can be regarded as realistic \cite{Kocha92,BildsC92}, 
we consider both states in order to exhibit the difference between them and
thus investigate the hydrodynamical effects on the main properties
of the binary system.

The fourth assumption regards the (zero-temperature) equation of state.
We choose a {\it polytropic} one:
\beq
  p=\kappa n^{\gamma},
\eeq
where $n$ is the fluid baryon number density,
and $\kappa$ and $\gamma$ are some constants.
In the present article, we consider that the two stars
in the binary system have the same $\kappa$ and $\gamma$,
even if they have different masses.

Finally, we assume that the spatial part of the metric
is conformally flat,
which corresponds to the {\em Isenberg-Wilson-Mathews (IWM)} approximation
of general relativity
\cite{Isenb78,IsenbN80,WilsoM89} (see
\cite{FriedUS02} for a discussion).
In this approximation, the spacetime metric takes the form:
\beq
  ds^2 =-(N^2 -B_i B^i) dt^2 -2B_i \, dt \, dx^i +A^2 f_{ij} dx^i dx^j,
  \label{eq:metric}
\eeq
$N$ being the lapse function, $B^i$ the shift vector of co-orbiting
coordinates, $A$ the conformal factor,
and $f_{ij}$ the flat spatial metric.
The accuracy of this approximation should be checked (see Sec.~III.A of
Paper~I).
The comparison between the IWM results presented here and
the non-conformally flat ones will be performed elsewhere \cite{Limou03}.

\subsection{Partial differential equations to be solved}

Under the assumptions listed above, we have derived the equations
to be integrated in Paper I. Here we present only some outline of them.

The gravitational field equations have been obtained within the 3+1 decomposition
of the Einstein's equation \cite{York79}.
The constraint equations are treated in the so-called 
conformal thin-sandwich approach
(see \cite{Cook00,BaumgS03} for a review), which enables to take
into account the helical symmetry of spacetime. 
The trace of the spatial part of the Einstein equation
combined with the Hamiltonian constraint results in two equations:
\beqa
  \ulap \nu &=& 4\pi A^2 (E+S) + A^2 K_{ij} K^{ij}
  - \unab_i \nu \unab^i \beta, \\
  \ulap \beta &=& 4\pi A^2 S +{3 \over 4} A^2 K_{ij} K^{ij}
  -{1 \over 2} (\unab_i \nu \unab^i \nu +\unab_i \beta \unab^i \beta),
\eeqa
where $\unab_i$ stands for the covariant derivative associated with
the flat 3-metric $f_{ij}$ and  $\ulap := \unab^i \unab_i$ for the 
associated Laplacian operator.
The quantities $\nu$ and $\beta$ are defined by
$\nu := \ln N$ and $\beta := \ln (AN)$, and
$K_{ij}$ denotes the extrinsic curvature tensor of the $t={\rm const}$
hypersurfaces.
$E$ and $S$ are respectively the matter energy density and the trace of the 
stress tensor, both as measured by the observer whose 4-velocity is the 
unit normal
$n^\mu$ to the $t={\rm const}$
hypersurfaces {\em (Eulerian observer)}:
\beqa
  E &:=& T_{\mu \nu} n^{\mu} n^{\nu} , \\
  S &:=& A^2 f_{ij} T^{ij}.
\eeqa
We have also to solve the momentum constraint, which writes
\beq
  \ulap N^i +{1 \over 3} \unab^i (\unab_j N^j) = -16\pi N A^2 (E+p) U^i
  +2 N A^2 K^{ij} \unab_j (3\beta -4\nu),
\eeq
where $N^i:=B^i +\Omega (\p/\p \vp)^i$ denotes the shift vector of
nonrotating coordinates.

Apart from the gravitational field equations,
we have to solve the fluid equations.
The equations governing the quasiequilibrium state are
the Euler equation and the equation of baryon number conservation.
Both cases of irrotational and synchronized motions admit a first
integral of the relativistic Euler equation (see Sec.~II of Paper~I),
which is written as
\beq
  H + \nu - \ln \Gamma_0 + \ln \Gamma = {\rm const.},
\eeq
where $H := \ln h$ is the logarithm of the fluid specific enthalpy:
\beq
  h := {e+p \over m_{\rm B} n},
\eeq
$m_{\rm B}$ being the mean baryon mass.
$\Gamma_0$ is the Lorentz factor between the co-orbiting
observer (we denote his 4-velocity by $v^{\mu}$) and the Eulerian observer:
\beq
  \Gamma_0 := -n^{\mu} v_{\mu},
\eeq
and $\Gamma$ is the Lorentz factor between the fluid and 
the co-orbiting observers:
\beq
  \Gamma := -v^{\mu} u_{\mu}.
\eeq
For a synchronized motion, the equation of baryon number conservation is
trivially satisfied, whereas for an irrotational flow, it is written as
\beq
  \zeta H \ulap \Psi +\unab^i H \unab_i \Psi
  = A^2 h \Gamma_{\rm n} U^i_0 \unab_i H
  + \zeta H [ \unab^i \Psi \unab_i (H-\beta)
    +A^2 h U^i_0 \unab_i \Gamma_{\rm n}],
\eeq
where $\Psi$ is the velocity potential and $\zeta$ the thermodynamical
coefficient:
\beq
  \zeta := {d \ln H \over d \ln n}.
\eeq
$\Gamma_{\rm n}$ denotes the Lorentz factor between the fluid and
the Eulerian observer:
\beq
  \Gamma_{\rm n} := -n^{\mu} u_{\mu},
\eeq
and $U^i_0$ is the orbital 3-velocity with
respect to the Eulerian observers:
\beq
  U^i_0 = -{B^i \over N}. \label{eq:ov_wrt_euler}
\eeq
Within the IWM approximation, the three Lorentz factors introduced 
so far have the following expressions:
\beqa
  \Gamma_0 &=& (1 -A^2 f_{ij} U^i_0 U^j_0)^{-1/2}, \\
  \Gamma_{\rm n} &=& (1 -A^2 f_{ij} U^i U^j)^{-1/2}, \\
  \Gamma &=& \Gamma_{\rm n} \Gamma_0 (1 -A^2 f_{ij} U^i U^j_0),
\eeqa
where $U^i$ is the fluid 3-velocity with respect to the Eulerian observer:
$U^i = U^i_0$ for synchronized binary systems, whereas 
\beq
  U^i = {1 \over A^2 \Gamma_{\rm n} h} \unab^i \Psi \label{eq:iv_wrt_corot}
\eeq
for irrotational ones.

\subsection{Numerical method}

Before going further, let us recall shortly
the numerical method that we employ, including   
the special technique to treat the case of
non-integer polytropic indices $n=1/(\gamma-1)$ \cite{BonazGM98,TanigGB01}.

Through this series of works, we have developed a numerical code
which is based on a multidomain spectral method
with surface-fitted coordinates \cite{BonazGM99b,GrandBGM01}.
This code is built upon the C++ library
LORENE \cite{lorene}.

One of the merits of considering multiple domains is
to allow for boundary conditions at infinity, by 
compactifying the outermost domain.
Another merit is
that we can separate the region including the matter from the other ones.
This property results in accurate computations,
by setting at the boundary between two domains the discontinuities 
which might occur at the stellar surface.
Nevertheless, some infinite gradient in the baryon density appears 
at the surface for stiff equations of state ($\gamma \geq 2$).
(Even for $\gamma<2$, if the polytropic index $n$ is not an integer,
higher derivatives of density diverge at the stellar surface).
In order to overcome this,
we have introduced a regularization technique in Ref.~\cite{BonazGM98}
and Paper II and we employ it in the present version of the code.


\begin{figure}[htb]
\vspace{7mm}
\centerline{\includegraphics[width=0.4\textwidth]{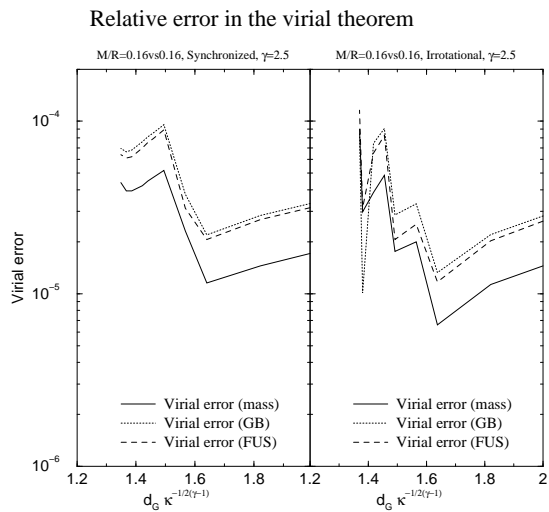}}
\caption[]{\label{fig:virial1}
Relative error in the virial theorem of stationary binary figures
in general relativity
as a function of the orbital separation.
Solid, dotted, and dashed curves correspond to relative errors in 
the three variants of the relativistic virial theorem (see text).
Left (right) panel is for synchronized (irrotational) binary systems.
Both of them are calculated for identical-mass binaries
with the adiabatic index $\gamma=2.5$ and the compactness
$M/R=0.16~{\rm vs}~0.16$.
}
\end{figure}%

\begin{figure}[htb]
\vspace{7mm}
\centerline{\includegraphics[width=0.4\textwidth]{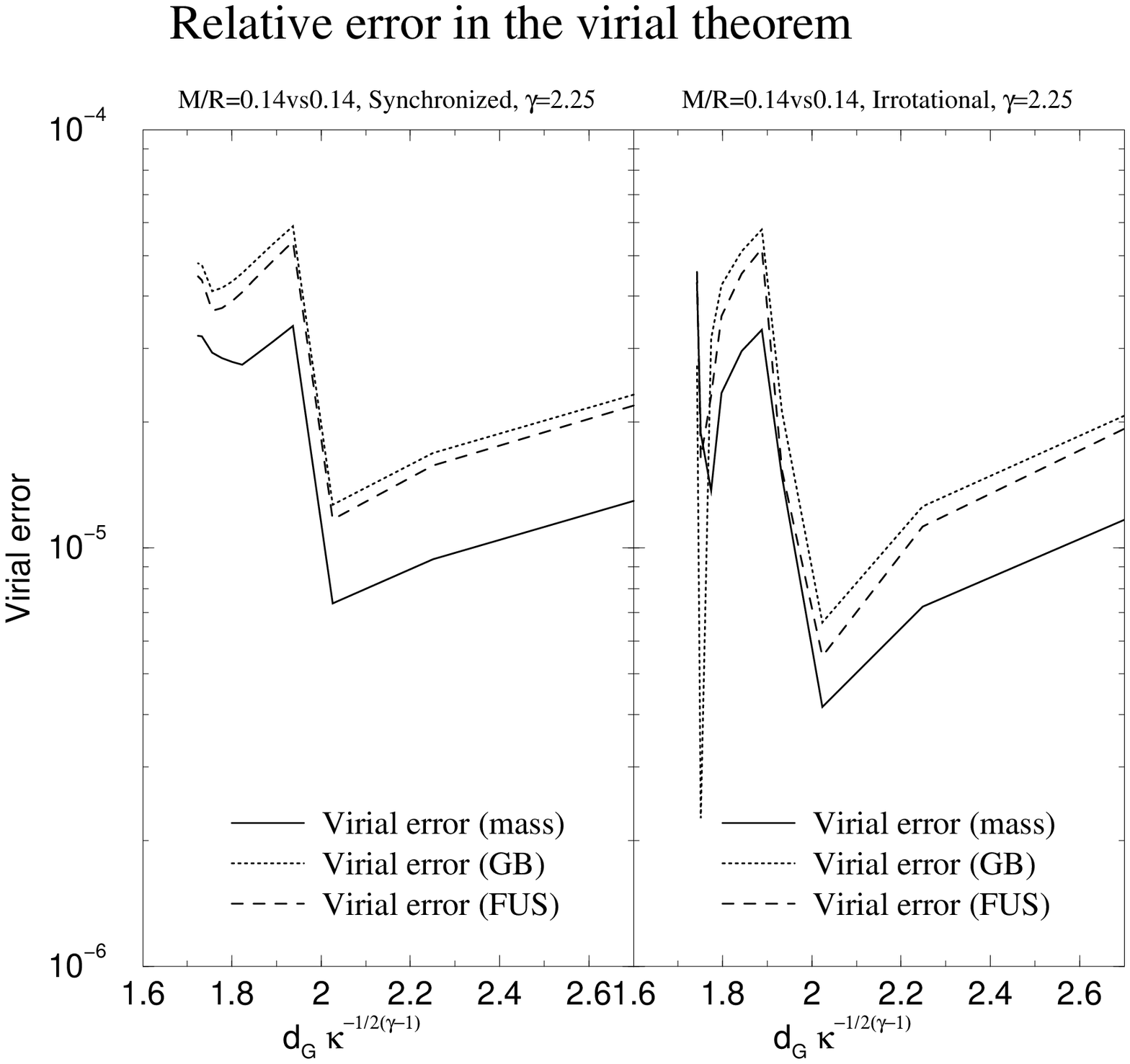}}
\caption[]{\label{fig:virial2}
Same as Fig. \ref{fig:virial1} but for $\gamma=2.25$ and
$M/R=0.14~{\rm vs}~0.14$.
}
\end{figure}%

\begin{figure}[htb]
\vspace{7mm}
\centerline{\includegraphics[width=0.4\textwidth]{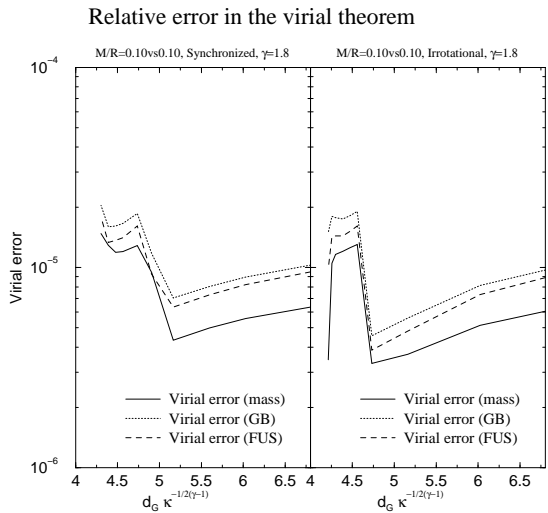}}
\caption[]{\label{fig:virial3}
Same as Fig. \ref{fig:virial1} but for $\gamma=1.8$ and
$M/R=0.10~{\rm vs}~0.10$.
}
\end{figure}%

\section{Code tests} \label{s:tests}

Numerous tests of the numerical code have already been presented in 
Papers~I, II, III and Ref.~\cite{TanigG02a}.
Consequently we presents here tests only regarding 
(i) the global numerical error measured by the virial theorem,
(ii) the numerical error in the first law of
binary neutron star thermodynamics,
(iii) the comparison with previous numerical solutions,
and (iv) the
convergence to post-Newtonian results at large separation. 

\subsection{Virial theorem}

The virial theorem has proved to be a useful tool to check the
global error in numerical solutions for stationary fluid systems
(see e.g. \cite{BonazGM98}).
A general relativistic version of the virial theorem has been
obtained by Gourgoulhon and Bonazzola \cite{GourgB94} for
stationary spacetimes. It has been recently extended to
binary star spacetimes within the IWM approximation by
Friedman, Uryu and Shibata \cite{FriedUS02}
\footnote{The spacetime generated by a binary system is
not stationary, due to gravitational radiation.
However, within the IWM approximation, the gravitational
radiation is neglected in the spacetime global dynamics,
so that one is able to recover the virial theorem.}.
The virial relation is equivalent to require
\beq
  M_{\rm ADM} = M_{\rm Komar},
\eeq
where $M_{\rm ADM}$ is the ADM mass:
\beq  \label{e:M_ADM}
  M_{\rm ADM} = -{1 \over 2\pi} \oint_{\infty} \unab^i A^{1/2} dS_i,
\eeq
and $M_{\rm Komar}$ is a Komar-type mass, defined by
\beq
  M_{\rm Komar} = {1 \over 4\pi} \oint_{\infty} \unab^i N dS_i.
\eeq
The virial theorem obtained by Friedman, Uryu and Shibata \cite{FriedUS02}
(see also Eq. (5.7) of Ref. \cite{ShibaU01}) writes
\beqa
  VE(FUS) &=&\int \Bigl[ 2N A^3 S +{3 \over 8\pi} N A^3 K_i^j K_j^i
	+{1 \over 4\pi} N A (\unab_i \beta \unab^i \beta
	-\unab_i \nu \unab^i \nu)
	\Bigr] \, d^3 x\nonumber \\
	&=&0. \label{eq:ve_fus}
\eeqa
By a straightforward manipulation, this integral can be recast in
the form of the virial theorem as obtained by Gourgoulhon and Bonazzola
\cite{GourgB94}:
\beqa
  VE(GB) &=&\int \Bigl[ 2A^3 S +{3 \over 8\pi} A^3 K_i^j K_j^i
	+{1 \over 4\pi} A (\unab_i \beta \unab^i \beta
	-\unab_i \nu \unab^i \nu
	-2 \unab_i \beta \unab^i \nu) \Bigr] \, d^3 x \nonumber \\
	&=&0 . \label{eq:ve_gb}
\eeqa
Let us stress that the above identity has been derived by
Gourgoulhon and Bonazzola \cite{GourgB94} only for stationary
spacetimes and that its validity for IWM spacetimes with
helical symmetry has been obtained by Friedman,
Uryu and Shibata \cite{FriedUS02}.

As error indicators for our numerical solutions,
we have evaluated the quantities
\beqa
  &&\Bigl| {M_{\rm ADM} - M_{\rm Komar} \over M_{\rm ADM}} \Bigr|,
  \label{eq:virial_mass} \\
  &&\Bigl| {VE(FUS) \over M_{\rm ADM}} \Bigr|, \label{eq:virial_fus} \\
  &&\Bigl| {VE(GB) \over M_{\rm ADM}} \Bigr|. \label{eq:virial_gb}
\eeqa
The results are presented in Figs. \ref{fig:virial1} -- \ref{fig:virial3}
as a function of the orbital separation $d_G$, defined by the distance
between the two ``center of mass'' (see Eq.~(128) of Paper~I for a precise
definition). 
In each figure, the left (resp. right) panel is for synchronized 
(resp. irrotational) binaries.
The vertical axis denotes the relative virial errors defined by
Eqs.~(\ref{eq:virial_mass})-(\ref{eq:virial_gb}).
One can notice in these panels 
a sudden change in the slope of the curves around the middle of the 
sequences.
This is due to the fact that we use five computational domains 
for each star and the space around it in the case of large separation
and four domains for small separation, inducing a change in the
accuracy of the computation.
However, since such discrete changes are only a factor two or three better
even if we use five computational domains for close separation,
we decided to use four domains in order to save computational time.
The order of magnitude of the virial error is $\sim 10^{-5}$
through the sequence.
This is sufficient to discuss the physics
of binary systems accurately.

\subsection{First law of binary neutron star thermodynamics}

The first law of binary neutron star thermodynamics within 
the IWM approximation
has been derived by Friedman, Uryu and Shibata \cite{FriedUS02}. 
It is written as
\beq \label{e:first_law}
  \delta M_{\rm ADM} = \Omega \, \delta J,
\eeq
where $\delta M_{\rm ADM}$ and $\delta J$ are respectively the changes
in ADM mass and total angular momentum along a constant baryon number sequence,
and $\Omega$ denotes the orbital angular velocity.
Since the relation (\ref{e:first_law}) is not enforced in our code, 
we may use it to gauge the numerical error. 
The relative error
$| (\delta M_{\rm ADM} - \Omega \delta J) / \delta M_{\rm ADM} |$
has been evaluated for some selected points in the sequences
by means of a second order polynomial fitting.
It is found that this error is smaller than a few
$10^{-3}$ for large separations
and $1 \times 10^{-2}$ for medium separations.
However, it becomes a few \% or worse for close configurations
and around the turning point ($\delta M_{\rm ADM}=0$) of the ADM mass.

\subsection{Comparison with previous numerical results}

We have compared our results with the irrotational sequences
computed by Uryu, Shibata and Eriguchi \cite{UryuSE00}.
Since only data at the sequence end points are listed in 
Ref.~\cite{UryuSE00},
we have compared for end point configurations only
and listed the results in Table~\ref{table0}. 
In the same table, 
results of Shibata and Uryu \cite{ShibaU01} 
are also given for comparison in the case $\gamma=2$.
It is found that the ADM mass agree with each other
at the order of $10^{-3}$, and that the orbital angular velocity 
as well as the total angular momentum agree at a few \%.
The orbital angular velocity of our results
is a few \% smaller than that of Uryu, Shibata and Eriguchi
because the end point of our sequences is still a few \% (in term
of $\Omega$) far from the mass-shedding limit, as discussed below
(Sec.~\ref{s:endpoint}).  
On the other hand, the difference in the total angular momentum $J$
seems to arise from some systematic error in Ref.~\cite{UryuSE00}.
Indeed our values for $J$ are smaller that of Ref.~\cite{UryuSE00}, 
although our binaries have a slightly larger separation.
Moreover, the new results given by them in Refs.~\cite{ShibaU01,Uryu03}
for the case of $\gamma=2$ agrees with ours at the order of $10^{-3}$,
while being slightly smaller than our result as expected.

\subsection{Comparison with point-mass post-Newtonian computations
at large separation}

This last type of test will be presented in Sec.~\ref{s:ADM_mass}
(see Figs.~\ref{fig:binding} -- \ref{fig:binding_detail}).


\begin{figure}[htb]
\centerline{\includegraphics[width=0.5\textwidth]{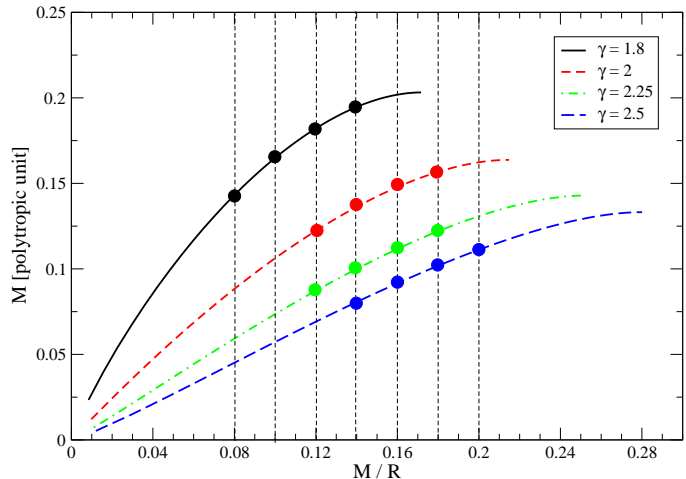}}
\caption[]{\label{fig:compact}
Single static neutron star models in the compactness - gravitational
mass plane. The right end of each curve is the maximum mass model.
The gravitational mass $M$ is rescaled by the polytropic constant $\kappa$:
$\bar M = \kappa^{-1/(2(\gamma-1))}\, M$. The dots correspond to 
values of compactness considered in the present work.
}
\end{figure}%


\begin{figure}[htb]
\vspace{7mm}
\centerline{\includegraphics[width=0.4\textwidth]{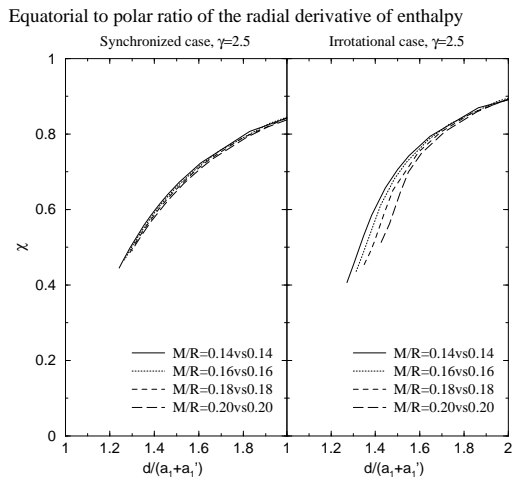}}
\caption[]{\label{fig:chi_eq_g25}
Equatorial to polar ratio of the radial derivative of the enthalpy 
(cusp indicator) as
a function of the orbital separation.
Left (right) panel is for synchronized (irrotational) binary systems
with the adiabatic index $\gamma=2.5$.
Solid, dotted, dashed, and long-dashed curves denote the cases
of compactness $M/R=0.14~{\rm vs}~0.14$, $0.16~{\rm vs}~0.16$,
$0.18~{\rm vs}~0.18$, and $0.20~{\rm vs}~0.20$, respectively.
}
\end{figure}%

\begin{figure}[htb]
\vspace{7mm}
\centerline{\includegraphics[width=0.4\textwidth]{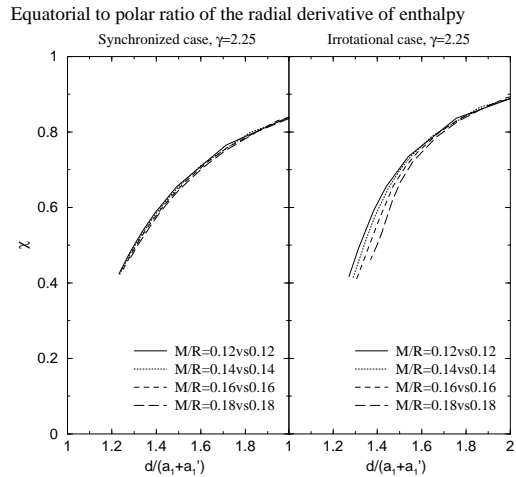}}
\caption[]{\label{fig:chi_eq_g225}
Same as Fig. \ref{fig:chi_eq_g25} but for $\gamma=2.25$.
Solid, dotted, dashed, and long-dashed curves denote the cases
of compactness $M/R=0.12~{\rm vs}~0.12$, $0.14~{\rm vs}~0.14$,
$0.16~{\rm vs}~0.16$, and $0.18~{\rm vs}~0.18$, respectively.
}
\end{figure}%

\begin{figure}[htb]
\vspace{7mm}
\centerline{\includegraphics[width=0.4\textwidth]{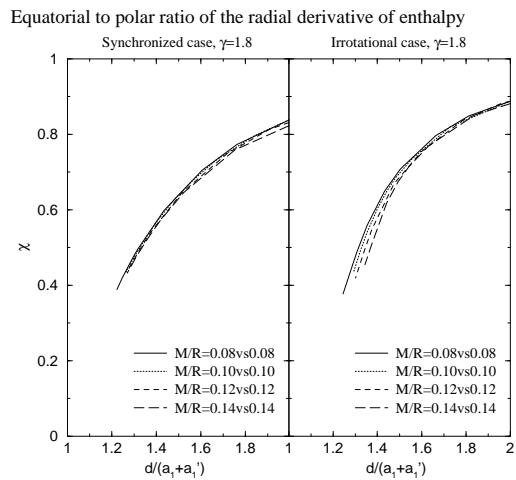}}
\caption[]{\label{fig:chi_eq_g18}
Same as Fig. \ref{fig:chi_eq_g25} but for $\gamma=1.8$.
Solid, dotted, dashed, and long-dashed curves denote the cases
of compactness $M/R=0.08~{\rm vs}~0.08$, $0.10~{\rm vs}~0.10$,
$0.12~{\rm vs}~0.12$, and $0.14~{\rm vs}~0.14$, respectively.
}
\end{figure}%

\begin{figure}[htb]
\vspace{7mm}
\centerline{\includegraphics[width=0.4\textwidth]{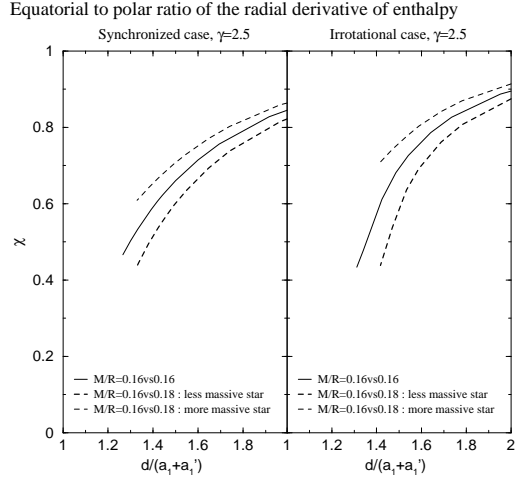}}
\caption[]{\label{fig:chi_df_g25}
Same as Fig. \ref{fig:chi_eq_g25} but for different-mass binary systems
compared with an identical-mass binary system.
The solid curve denotes the case of identical-mass binary system with
compactness $M/R=0.16~{\rm vs}~0.16$.
Thin and thick dashed curves are, respectively, for the more massive
and less massive star of a different-mass binary system
with $M/R=0.16~{\rm vs}~0.18$.
}
\end{figure}%

\begin{figure}[htb]
\vspace{7mm}
\centerline{\includegraphics[width=0.4\textwidth]{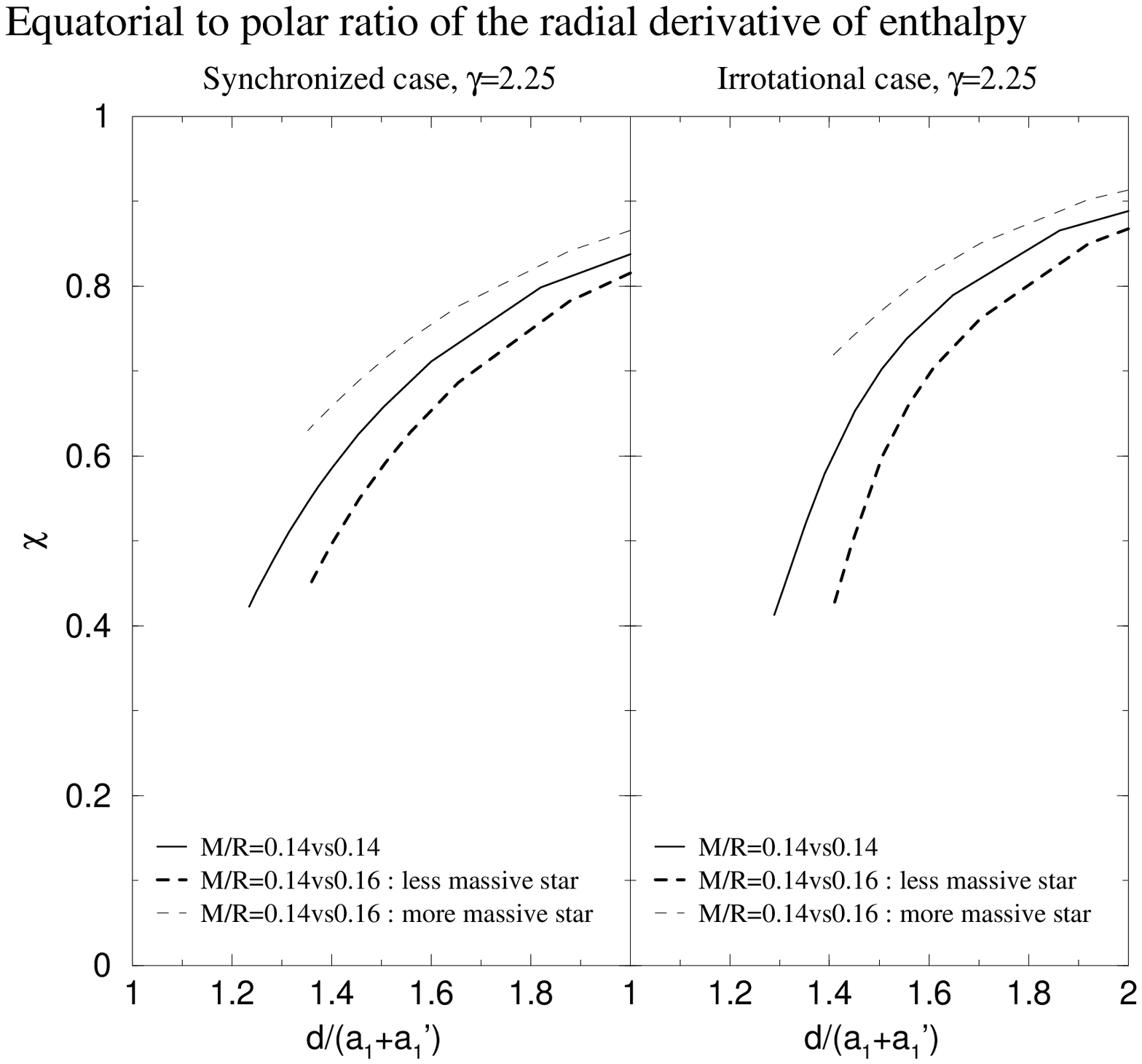}}
\caption[]{\label{fig:chi_df_g225}
Same as Fig. \ref{fig:chi_df_g25} but for $\gamma=2.25$
The solid curve denotes the case of identical-mass binary system with
compactness $M/R=0.14~{\rm vs}~0.14$.
Thin and thick dashed curves are, respectively, for the more massive
and less massive star of a different-mass binary system
with $M/R=0.14~{\rm vs}~0.16$.
}
\end{figure}%

\begin{figure}[htb]
\vspace{7mm}
\centerline{\includegraphics[width=0.4\textwidth]{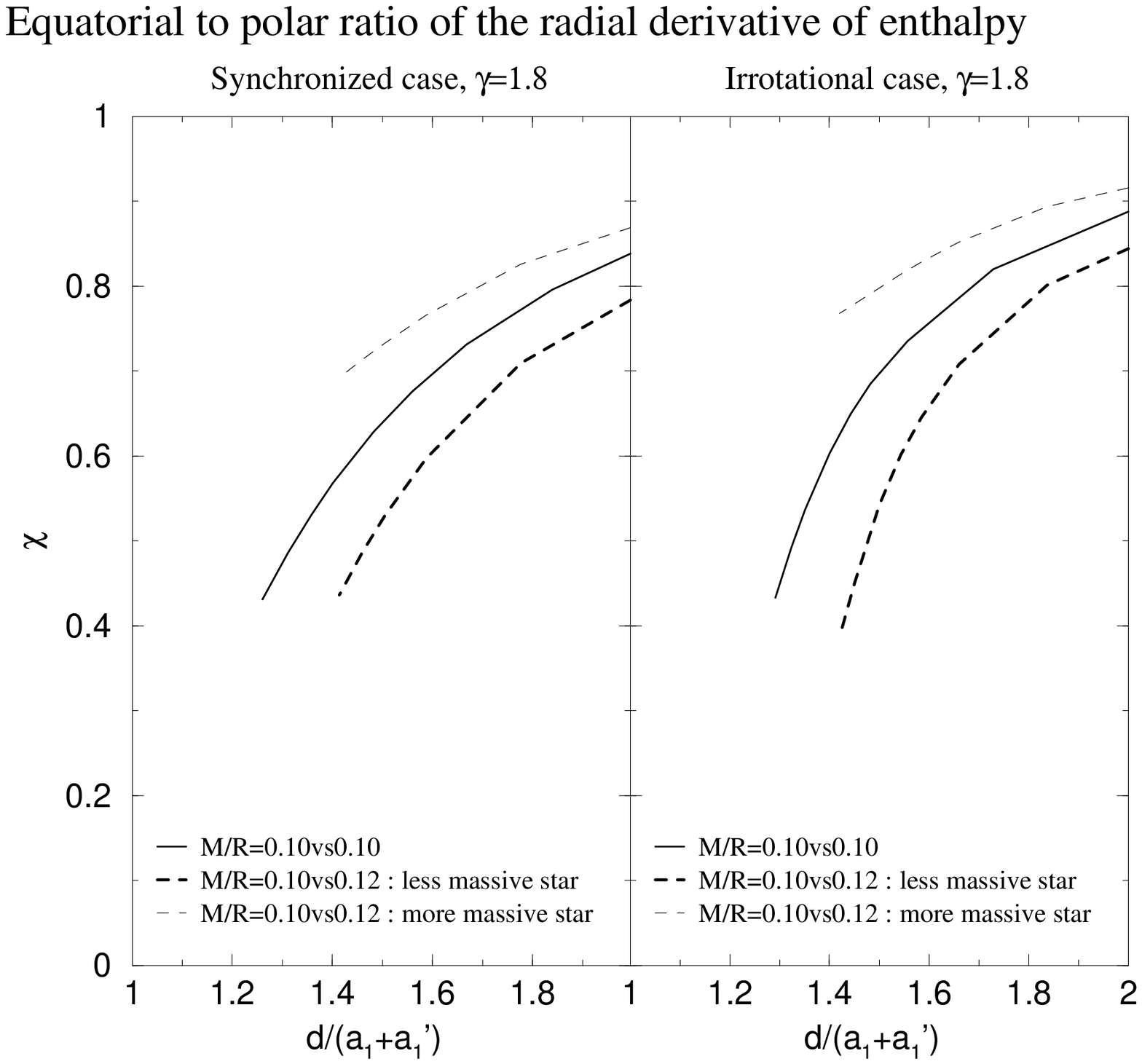}}
\caption[]{\label{fig:chi_df_g18}
Same as Fig. \ref{fig:chi_df_g25} but for $\gamma=1.8$
The solid curve denotes the case of identical-mass binary system with
compactness $M/R=0.10~{\rm vs}~0.10$.
Thin and thick dashed curves are,  respectively for the more massive
and less massive star of a different-mass binary system
with $M/R=0.10~{\rm vs}~0.12$.
}
\end{figure}%

\begin{figure}[htb]
\vspace{7mm}
\centerline{\includegraphics[width=0.4\textwidth]{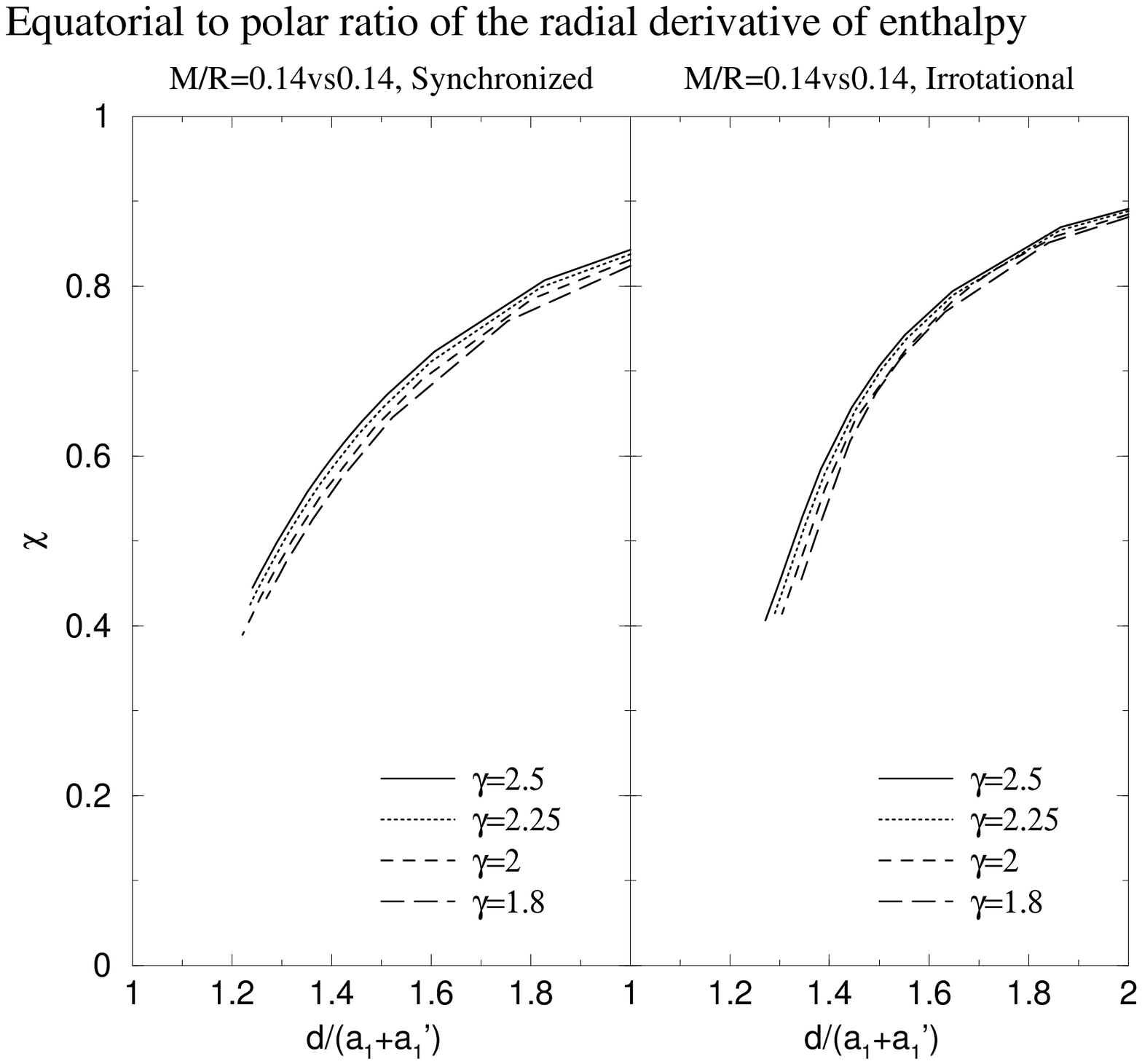}}
\caption[]{\label{fig:chi_gamma}
Equatorial to polar ratio of the radial derivative of the enthalpy 
(cusp indicator) as
a function of the orbital separation.
Left (right) panel is for synchronized (irrotational) binary systems
with the compactness $M/R=0.14~{\rm vs}~0.14$.
Solid, dotted, dashed, and long-dashed curves denote the cases
of adiabatic indices $\gamma=2.5$, 2.25, 2, and 1.8, respectively.
}
\end{figure}%

\section{Numerical results} \label{s:results}

The spirit of the quasiequilibrium approach is to model the 
(gravitational-radiation driven) evolution of a neutron star binary
by a sequence of steady-state circular-orbit configurations.
Along the sequence the baryon number is kept constant, since
no matter loss occurs during the evolution of each star, until
the end of the inspiral stage. Such constant-baryon-number sequences
are called {\em evolutionary sequences}.    

We have computed numerous evolutionary sequences, for different 
values of the adiabatic index $\gamma$ and various compactness.
We have used 5 (resp. 4) 
domains for each star and the space around it in the case of 
a large (resp. small) separation. In each domain, the 
number of collocation points of the spectral method
is chosen to be 
$N_r \times N_{\theta} \times N_{\vp} = 25 \times 17 \times 16$
or $33 \times 25 \times 24$,
where $N_r$, $N_{\theta}$, and $N_{\vp}$ denote
the number of collocation points in respectively the radial, polar,
and azimuthal directions.

As in Paper III,
we parametrize the difference in baryon mass
between the two stars by the values of compactness
of the stars like 
$[(M/R)_{\infty, \rm star~1}$ vs $(M/R)_{\infty, \rm star~2}]$,
where $(M/R)_{\infty}$ denotes the compactness of the considered
star at infinite
separation or equivalently that of an isolated spherical
star with the same baryon number.
Hereafter, we abbreviate $(M/R)_{\infty, \rm star~1}$
vs $(M/R)_{\infty, \rm star~2}$ as $M/R$ for simplicity.
According to recent nuclear equations of state,
the compactness of a $M=1.4 \, M_\odot$ neutron star is in the range
$0.12 \leq M/R \leq 0.20$, depending on the
equation of state \cite{Haens03}. We therefore explore this range
of compactness for our polytropic equations of state. 
We display the compactness parameters used for each adiabatic index $\gamma$
in Fig.~\ref{fig:compact}. For each couple of values $(\gamma,M/R)$,
one can see in Fig.~\ref{fig:compact} how far the considered model
is from the maximum mass configuration.

Since we have already shown numerous results for $\gamma=2$
in Paper III,
we focus here on results for $\gamma=2.5$, $2.25$, and $1.8$.
We also consider the case $\gamma=3$
in Sec. \ref{s:discussion}. However note that this value is too large
for a realistic equation of state of neutron star.

\subsection{End points of the sequences} \label{s:endpoint}

The start of mass exchange between the two stars marks the end
of the existence of quasiequilibrium configurations. This 
mass-shedding corresponds to the appearance of a cusp
at the surface of the lightest star (or both stars in the equal-mass
case). This cusp reveals the vanishing of the gradient of enthalpy
in the direction of the companion. Hence we consider 
the equatorial to polar ratio of the radial derivative
of the enthalpy 
\beq
  \chi := {(\p H/\p r)_{\rm eq, comp} \over (\p H/\p r)_{\rm pole}}
\eeq
as a dimensionless cusp indicator: $\chi=1$ at infinite separation, whereas 
$\chi=0$ when the cusp appears. The variation of $\chi$ along
various evolutionary sequences is shown
in Figs. \ref{fig:chi_eq_g25} -- \ref{fig:chi_eq_g18}
for identical-mass binary systems and
in Figs. \ref{fig:chi_df_g25} -- \ref{fig:chi_df_g18}
for different-mass systems compared with identical-mass ones.
The horizontal axis presents the orbital coordinate separation between
the two coordinate centers (maxima of density), $d$,
normalized by the summation of the coordinate radii of stars in the direction
of their companions ($a_1$ and $a_1'$)\footnote{In Paper III, Eq.~(15)
should be replaced by this equation.}:
\beq \label{e:def_tilde_d}
	{\tilde d} := \frac{d}{a_1+a_1'} .
\eeq
This dimensionless quantity is
an indicator of the contact between the two stars: 
${\tilde d} = 1$ corresponds
to the two stellar surfaces touching each other. 
Note here that $d$, $a_1$, and $a_1'$ are all {\em coordinate} lengths,
our coordinates being defined by the line element (\ref{eq:metric}).

{}From Figs. \ref{fig:chi_eq_g25} -- \ref{fig:chi_eq_g18},
one can see that the lines seem to focus toward the point $d/(a_1+a_1')=1$
and $\chi=0$ in the case of synchronized binary systems.
This implies that synchronized binary systems of identical-mass stars
in quasiequilibrium end their sequences by the contact between the two stars.
It is worth to point out that the curves for all cases of compactness
take almost the same tracks,
but very slightly deviate from each other:
the curves of larger compactness are located below those of smaller 
compactness.
On the other hand, in the case of irrotational binary systems,
the curves seem to reach $\chi=0$ at ${\tilde d} >1$.
This means that the end point of quasiequilibrium sequences
of irrotational binary systems is a {\em detached}
configuration with a cusp point. 
Let us recall that in Paper~II, we have found the same result
in the Newtonian regime. 
This tendency is enhanced for the case of larger compactness.
Note here that in Figs. \ref{fig:chi_eq_g25} and \ref{fig:chi_eq_g225},
the curves with larger compactness in the case of irrotational binary systems
turn slightly toward the vertical axis  around $\chi \sim 0.5$.
This comes from the numerical error.

The quasiequilibrium sequences should exist up to $\chi=0$.
However, the multidomain spectral method we employ cannot handle 
cusp-like figures because in the surface-fitted coordinate procedure
we assume that the stellar surface is smooth (see Sec.~IV.E of Paper~I
for a detailed discussion of this point). 
Indeed our numerical scheme fails to converge when $\chi\rightarrow 0$.
Monitoring the convergence by the relative difference $\d H$ 
between the enthalpy fields of two successive computational steps,
we hardly reach $\d H < 10^{-5}$ for synchronized binary systems
when the orbital separation is such that $\chi \sim 0.4$.
For irrotational configurations,
although it is still possible to reach $\d H \sim 10^{-6}$
when the orbital separation is such that $\chi \sim 0.4 - 0.5$,
small but unphysical oscillations of the stellar surface appear.
This results in the spurious (albeit slight) change of direction near the 
end of some curves in Figs.~\ref{fig:chi_eq_g25} and \ref{fig:chi_eq_g225}. 
These tendencies are enhanced for stiffer equation of state
(larger values of $\gamma$) and more compact stars (larger values of $M/R$).
Therefore, we stop the computations around $\chi \sim 0.4 - 0.5$.

We compare the evolution of the cusp indicator $\chi$ for 
different-mass binaries with
that for identical-mass ones in Figs. \ref{fig:chi_df_g25}
-- \ref{fig:chi_df_g18}. When extrapolating the curves toward $\chi=0$,  
the sequences seem to terminate
by a cusp at the surface of the lightest star, both for synchronized
and irrotational systems, while the heaviest star keeps its shape 
close to a spherical one.
It is worth to note here that as soon as the two masses differ,
even slightly, the sequence of a synchronized binary system
terminates by a detached mass-shedding limit (Roche limit).
Indeed the line for the less massive star is always located
below that for identical-mass stars which reach $\chi=0$
at ${\tilde d}=1$.
This implies that the line for the less massive star seems to always
reach $\chi=0$ at ${\tilde d} > 1$.

In Fig. \ref{fig:chi_gamma}, the evolution of $\chi$ is shown
for a fixed value of the compactness parameter,
$M/R=0.14~{\rm vs}~0.14$, and various values of the adiabatic index
$\gamma$.
It appears clearly that the curves of small $\gamma$
are located below those of large $\gamma$.
This means that at fixed relative orbital separation ${\tilde d}$ and
fixed compactness, a configuration with small $\gamma$ is closer to
the mass-shedding limit than a configuration with large
$\gamma$.
The reason is that stars with small adiabatic index have centrally
concentrated structures with some extended low density halos.
Such low density outer layers result in a smaller $\chi$.


\begin{figure}[htb]
\vspace{7mm}
\centerline{\includegraphics[width=0.4\textwidth]{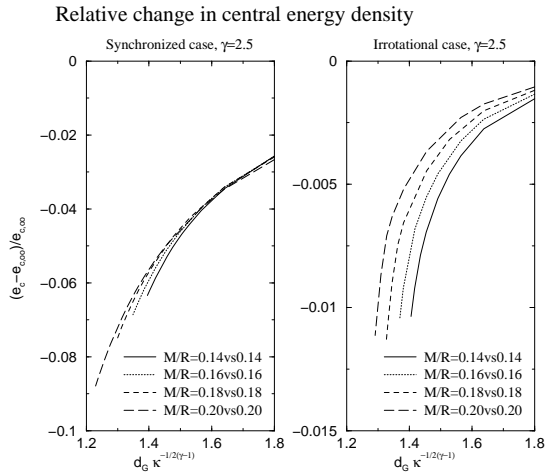}}
\caption[]{\label{fig:ec_eq_g25}
Relative change in central energy density as a function of the orbital
separation between the centers of mass of each star.
Left (right) panel is for synchronized (irrotational) binary systems
with the adiabatic index $\gamma=2.5$.
Solid, dotted, dashed, and long-dashed curves denote 
identical-mass binary systems with compactness
$M/R=0.14~{\rm vs}~0.14$, 0.16 vs 0.16, 0.18 vs 0.18, and
0.20 vs 0.20, respectively.
}
\end{figure}%

\begin{figure}[htb]
\vspace{7mm}
\centerline{\includegraphics[width=0.4\textwidth]{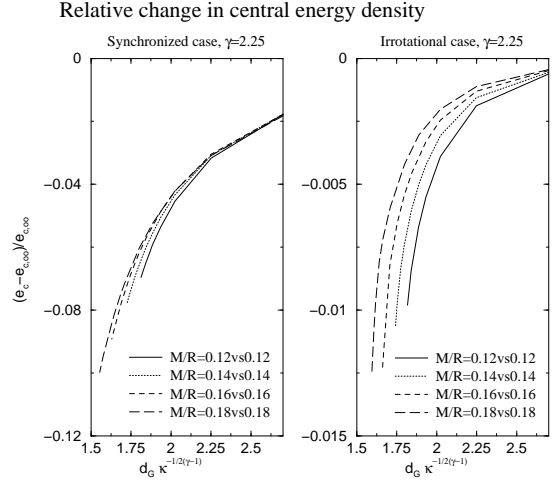}}
\caption[]{\label{fig:ec_eq_g225}
Same as Fig. \ref{fig:ec_eq_g25} but for $\gamma=2.25$.
Solid, dotted, dashed, and long-dashed curves denote 
identical-mass binary systems with compactness
$M/R=0.12~{\rm vs}~0.12$, 0.14 vs 0.14, 0.16 vs 0.16, and
0.18 vs 0.18, respectively.
}
\end{figure}%

\begin{figure}[htb]
\vspace{7mm}
\centerline{\includegraphics[width=0.4\textwidth]{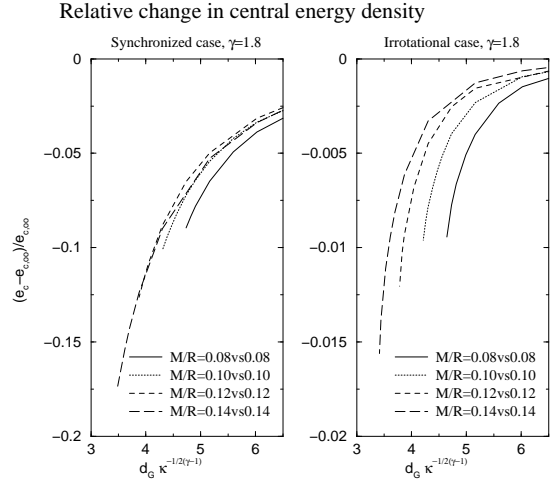}}
\caption[]{\label{fig:ec_eq_g18}
Same as Fig. \ref{fig:ec_eq_g25} but for $\gamma=1.8$
Solid, dotted, dashed, and long-dashed curves denote 
identical-mass binary systems with compactness
$M/R=0.08~{\rm vs}~0.08$, 0.10 vs 0.10, 0.12 vs 0.12, and
0.14 vs 0.14, respectively.
}
\end{figure}%

\begin{figure}[htb]
\vspace{7mm}
\centerline{\includegraphics[width=0.4\textwidth]{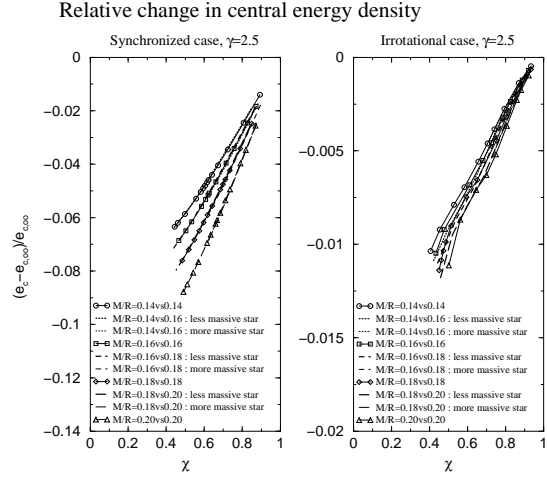}}
\caption[]{\label{fig:ec-chi_g25}
Relative change in central energy density as a function of the cusp
indicator $\chi$.
Left (right) panel is for synchronized (irrotational) binary systems
with the adiabatic index $\gamma=2.5$.
Solid curves with open circle, square, diamond, and triangle denote
identical-mass binary systems of compactness
$M/R=0.14~{\rm vs}~0.14$, 0.16 vs 0.16, 0.18 vs 0.18, and 0.20 vs 0.20,
respectively.
Thick and thin dotted curves are for the less and more massive stars
of the binary system with $M/R=0.14~{\rm vs}~0.16$,
thick and thin dashed curves for those with $M/R=0.16~{\rm vs}~0.18$,
and thick and thin long-dashed curves for those with $M/R=0.18~{\rm vs}~0.20$.
}
\end{figure}%

\begin{figure}[htb]
\vspace{7mm}
\centerline{\includegraphics[width=0.4\textwidth]{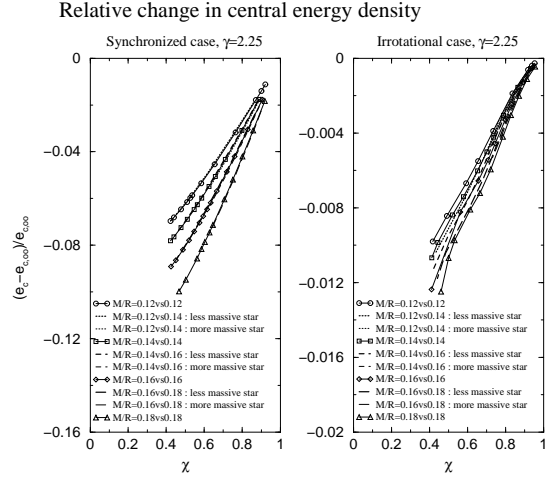}}
\caption[]{\label{fig:ec-chi_g225}
Same as Fig. \ref{fig:ec-chi_g25} but for $\gamma=2.25$.
Solid curves with open circle, square, diamond, and triangle denote
identical-mass binary systems of compactness
$M/R=0.12~{\rm vs}~0.12$, 0.14 vs 0.14, 0.16 vs 0.16, and 0.18 vs 0.18,
respectively.
Thick and thin dotted curves correspond respectively to 
the less and more massive stars
of the binary system with $M/R=0.12~{\rm vs}~0.14$,
thick and thin dashed curves for those with $M/R=0.14~{\rm vs}~0.16$,
and thick and thin long-dashed curves for those with $M/R=0.16~{\rm vs}~0.18$.
}
\end{figure}%

\begin{figure}[htb]
\vspace{7mm}
\centerline{\includegraphics[width=0.4\textwidth]{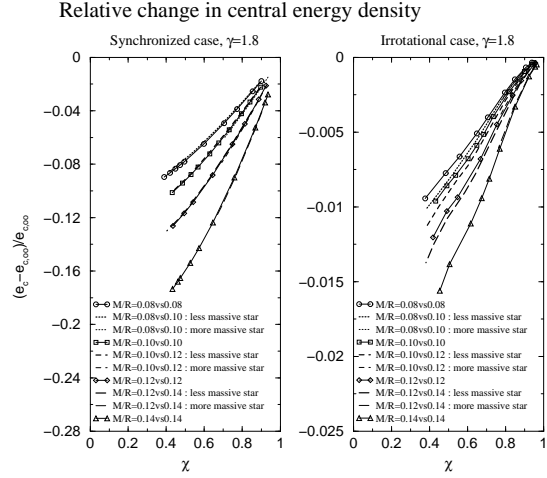}}
\caption[]{\label{fig:ec-chi_g18}
Same as Fig. \ref{fig:ec-chi_g25} but for $\gamma=1.8$.
Solid curves with open circle, square, diamond, and triangle denote
identical-mass binary systems of compactness
$M/R=0.08~{\rm vs}~0.08$, 0.10 vs 0.10, 0.12 vs 0.12, and 0.14 vs 0.14,
respectively.
Thick and thin dotted curves are for the less and more massive stars
of the binary system with $M/R=0.08~{\rm vs}~0.10$,
thick and thin dashed curves for those with $M/R=0.10~{\rm vs}~0.12$,
and thick and thin long-dashed curves for those with $M/R=0.12~{\rm vs}~0.14$.
}
\end{figure}%

\begin{figure}[htb]
\vspace{7mm}
\centerline{\includegraphics[width=0.4\textwidth]{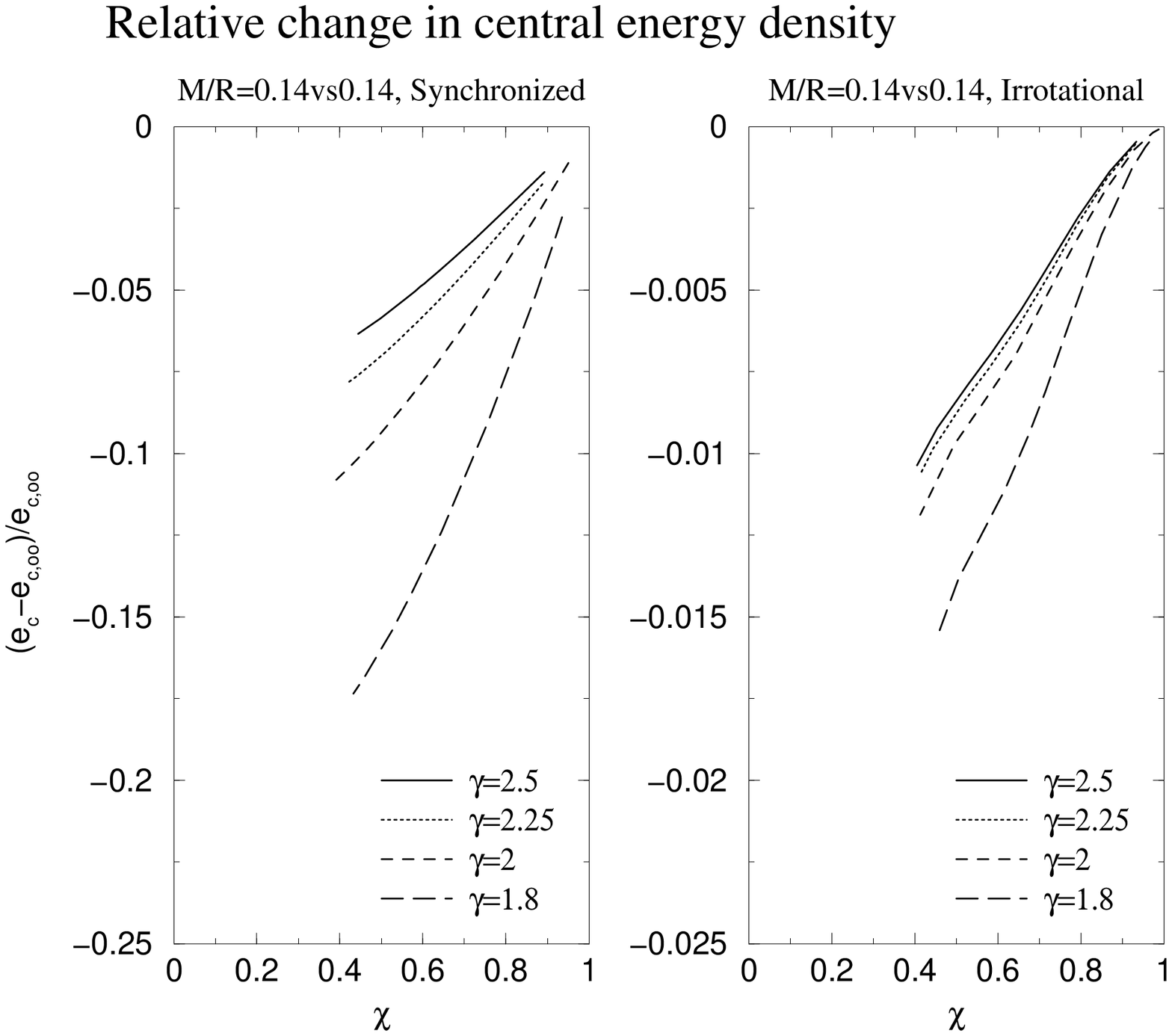}}
\caption[]{\label{fig:ec_gamma}
Relative change in central energy density as a function of the cusp
indicator $\chi$.
Left (right) panel is for synchronized (irrotational) binary systems
with compactness $M/R=0.14~{\rm vs}~0.14$.
Solid, dotted, dashed, and long-dashed curves denote the cases of
$\gamma=2.5$, 2.25, 2, and 1.8, respectively.
}
\end{figure}%

\subsection{Relative change in central density}

Let us now examine the relative change in central energy density
along an evolutionary sequence. 
In Figs. \ref{fig:ec_eq_g25} -- \ref{fig:ec_eq_g18} in depicted the
variation of the quantity
\beq \label{e:rel_e_c}
  \d e_c := {e_c - e_{c, \infty} \over e_{c, \infty}},
\eeq
where $e_c$ and $e_{c, \infty}$ denote the central proper energy density
of respectively the actual star and an isolated spherical star with 
the same baryon mass
(or in other words, that at infinite separation).
One can see from these figures that for all sequences the
central energy density decreases as the orbit shrinks.
It is also found that this decrease
is larger for larger compactness, at fixed adiabatic index and 
rotation state.
Note that the magnitude of the decrease is very
different between the synchronized case and the irrotational one:
$\sim 10\%
$ for synchronized binaries versus a few $\%
$
for irrotational ones. 

The relative change in central energy density is
presented as a function of the cusp indicator $\chi$
in Figs. \ref{fig:ec-chi_g25} -- \ref{fig:ec-chi_g18}, in order
to evaluate the total change at the end of the sequence
($\chi=0$). 
It is found from these figures that the central energy density
for large compactness
decreases more than that for small compactness.
Furthermore, we can see that the decreasing track of central energy density
is determined by the compactness of the star itself and
not by that of its companion,
even though the value itself is affected by the companion star.
This effect is clearly found for synchronized binary systems.
For irrotational systems, 
since the amount of decrease is very small and the curves include some
numerical errors,
it is not clear whether the decrease depends only on the compactness
of the star and not on that of its companion. However the same tendency 
could be found, in particular in the case of
a soft equation of state like $\gamma=1.8$ (see Fig. \ref{fig:ec-chi_g18}).
Note here that the slight changes of direction in the curves 
around $\chi \sim 0.5$ for large compactness in the irrotational case in Figs.
\ref{fig:ec-chi_g25} and \ref{fig:ec-chi_g225} is due to the  
numerical error discussed in the previous subsection \ref{s:endpoint}.

Although the sequences stop at around $\chi \simeq 0.4$,
it is possible to predict how much the central energy density decreases
at the mass-shedding limit by extrapolating the curves
in Figs. \ref{fig:ec-chi_g25} -- \ref{fig:ec-chi_g18}.
For instance, the central energy density of
synchronized binary systems with $\gamma=2.25$ may decrease
by 10 \% -- 15 \% in the range of $M/R=0.12~{\rm vs}~0.12$ --
0.18 vs 0.18.

Figure \ref{fig:ec_gamma} is shown
in order to investigate the behavior of the relative change in central
energy density when varying the adiabatic index and keeping the compactness
fixed. 
It is found that the rate of decrease is lower for larger adiabatic indices.
This is because the larger the adiabatic index, the closer the
matter to the incompressible state in which the central energy density
does not change.


\begin{figure}[htb]
\vspace{7mm}
\centerline{\includegraphics[width=0.5\textwidth]{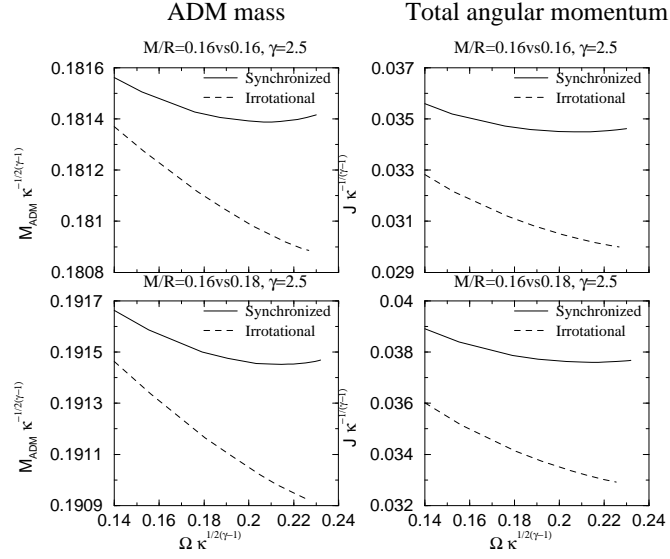}}
\caption[]{\label{fig:admang_g25}
ADM mass and total angular momentum of the binary system
as a function of the orbital angular velocity.
Top (bottom) panels are for identical-mass (different-mass)
binary systems of compactness $M/R=0.16~{\rm vs}~0.16$
($M/R=0.16~{\rm vs}~0.18$) with the adiabatic index $\gamma=2.5$.
In each panel, the solid curve denotes the synchronized case,
and the dashed one the irrotational one.
}
\end{figure}%

\begin{figure}[htb]
\vspace{7mm}
\centerline{\includegraphics[width=0.5\textwidth]{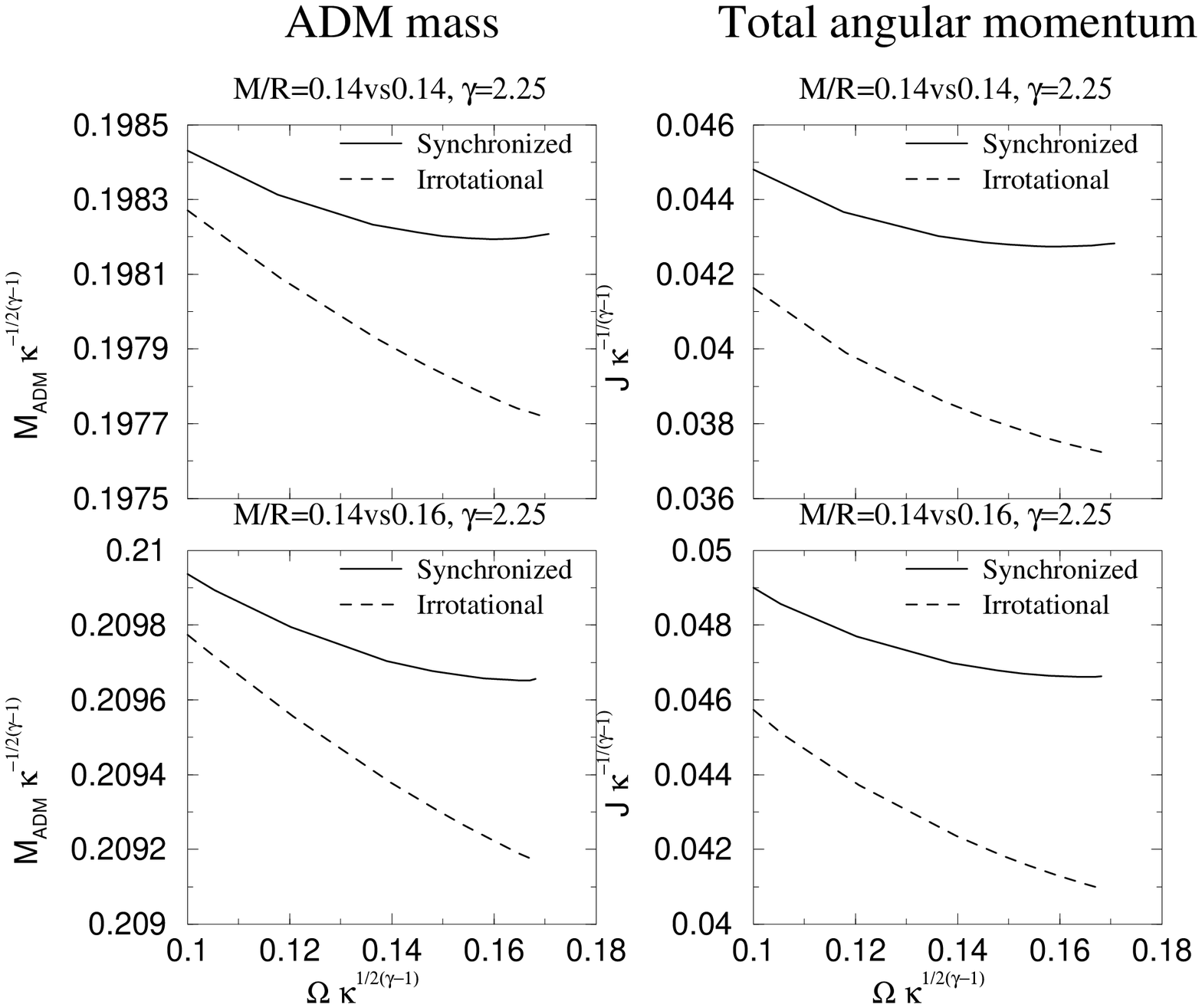}}
\caption[]{\label{fig:admang_g225}
Same as Fig. \ref{fig:admang_g25} but for $\gamma=2.25$.
Top (bottom) panels are for identical-mass (different-mass) binary systems
of compactness $M/R=0.14~{\rm vs}~0.14$ ($M/R=0.14~{\rm vs}~0.16$).
}
\end{figure}%

\begin{figure}[htb]
\vspace{7mm}
\centerline{\includegraphics[width=0.5\textwidth]{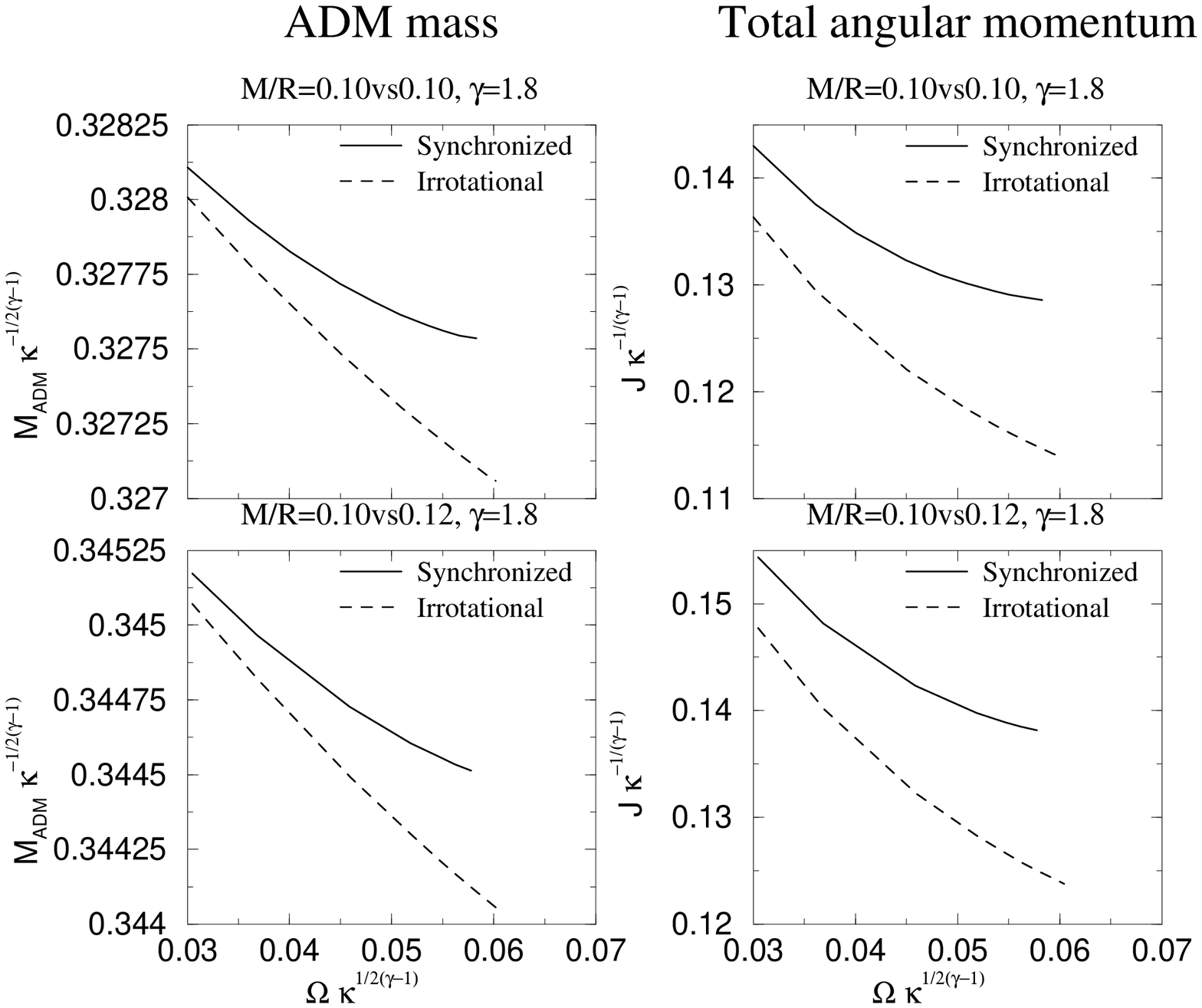}}
\caption[]{\label{fig:admang_g18}
Same as Fig. \ref{fig:admang_g25} but for $\gamma=1.8$.
Top (bottom) panels are for identical-mass (different-mass) binary systems
of compactness $M/R=0.10~{\rm vs}~0.10$ ($M/R=0.10~{\rm vs}~0.12$).
}
\end{figure}%


\begin{figure}[htb]
\vspace{7mm}
\centerline{\includegraphics[width=0.6\textwidth]{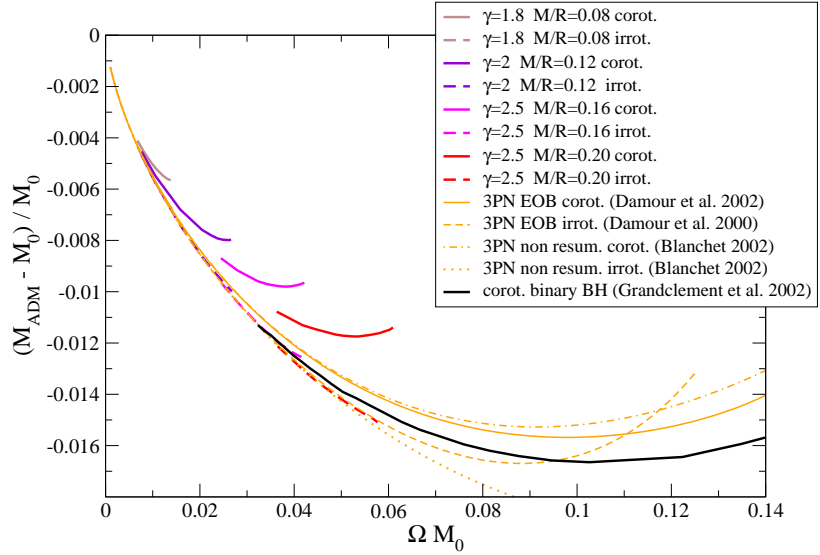}}
\caption[]{\label{fig:binding}
Relative binding energy along evolutionary sequences of equal-mass 
binary neutron stars,
compared with (i) analytical results at the 3rd post-Newtonian order
for point-masses by Damour et al. 2000  \cite{DamouJS00},
Damour et al. 2002 \cite{DamouGG02} and Blanchet 2002 \cite{Blanc02a},
and with (ii) numerical results for corotating binary black holes
by Grandcl\'ement et al. 2002 \cite{GrandGB02}. $\Omega$ is the 
orbital angular velocity and $M_0$ is twice the gravitational
mass of a single static neutron star (resp. black hole) of the same 
baryon number (resp. same horizon area) as that defining the 
considered sequence. 
}
\end{figure}%

\begin{figure}[htb]
\vspace{7mm}
\centerline{\includegraphics[width=0.6\textwidth]{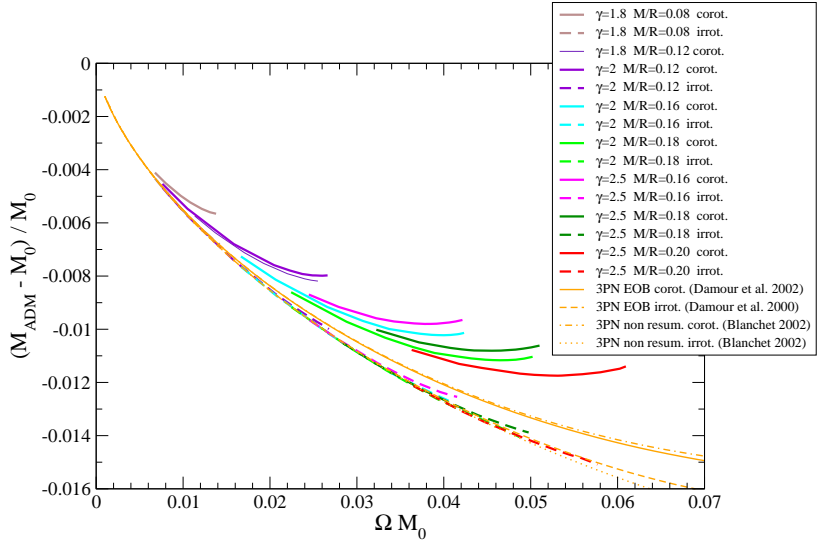}}
\caption[]{\label{fig:binding_detail}
Enlargement of Fig.~\ref{fig:binding} in the neutron star
region, with some additional sequences.
}
\end{figure}%

\subsection{ADM mass and total angular momentum} \label{s:ADM_mass}

The variation of the ADM mass $M_{\rm ADM}$ and total angular momentum
$J$ 
along an evolutionary sequence is  presented
in Figs. \ref{fig:admang_g25} -- \ref{fig:admang_g18}.
These figures show that a {\em turning point}
(minimum of $M_{\rm ADM}$ or $J$) appears for
synchronized binaries with $\gamma=2.5$ and $2.25$.
On the other hand, we cannot find it in the synchronized case
with $\gamma=1.8$ and in any of irrotational sequences.
Note that the position along the sequence of the turning point
in the ADM mass and the total angular momentum 
coincide with each other, as expected from
Eq.~(\ref{e:first_law}).
The turning points are also found in the different-mass case
for synchronized binary systems with $\gamma=2.5$ and 2.25.
However, their position is closer to the end point of the sequence
than that in the identical-mass case.
Since, as mentioned above, we stop the computation slightly 
before the end point  of quasiequilibrium sequences,
there remains the possibility that the turning point may exist
in the range $0< \chi < 0.4$,
even if we do not find it in the synchronized case with $\gamma=1.8$
and in the irrotational case.
The complete discussion about the existence of the turning point
for various adiabatic indices and compactness will be presented 
in Sec. \ref{s:discussion}.

In Figs.~\ref{fig:binding} -- \ref{fig:binding_detail}, we present
the ADM mass as a function of the orbital angular velocity,
as in Figs.~\ref{fig:admang_g25} -- \ref{fig:admang_g18}, but 
with a scaling defined by twice the gravitational mass of
the single static neutron star of the same baryon number as that
of the sequence, $M_0$ (whereas
in Figs.~\ref{fig:admang_g25} -- \ref{fig:admang_g18},
the scaling was set by the polytropic constant $\kappa$).
Note that $M_0$ can also be viewed as the ADM mass at infinite separation. 
This enables us to compare with third order post-Newtonian results 
for point-mass particles obtained in the Effective One
Body (EOB) approach by Damour et al. \cite{DamouJS00,DamouGG02}
or in the standard non-resummed post-Newtonian framework by Blanchet
\cite{Blanc02a}. This also enables us to compare with corotating 
binary black holes, according to the numerical results by
Grandcl\'ement et al. \cite{GrandGB02}. We note from these figures
that the irrotational binary neutron star sequences are very close
to the 3PN irrotational ones (dashed and dotted fine curves).
Moreover, all curves converges at large separation (small value of
$\Omega M_0$), which can be considered as a check of our numerical
computations, especially the way $\Omega$ is determined
(cf. Sec.~IV.D.2 of Paper~I). 

Another interesting feature shown in Fig.~\ref{fig:binding}
is the good agreement between the irrotational binary neutron star sequence
with high compactness ($M/R=0.20$) and the binary black hole sequence. 
The latter is made of corotating black holes, but for $\Omega M_0 \leq 0.06$,
the spin effects are not very important, so that it is meaningful to 
compare the corotating black hole sequence up to $\Omega M_0 \leq 0.06$
with the irrotational neutron star sequence. 
Finally, we notice from Fig.~\ref{fig:binding_detail} that at
a given compactness and separation, the configuration with smaller 
adiabatic index
is more bound. Conversely, at a given adiabatic index and separation, the 
configuration with higher compactness is more bound. 


\begin{figure}[htb]
\vspace{7mm}
\centerline{\includegraphics[width=0.4\textwidth]{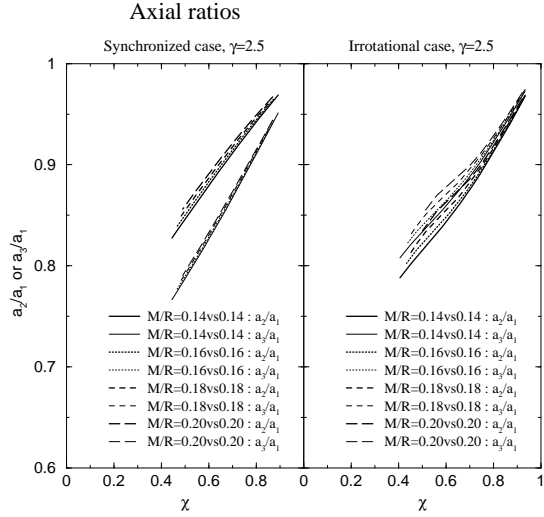}}
\caption[]{\label{fig:axis_eq_g25}
Axial ratios $a_2/a_1$ (thick curves) or $a_3/a_1$ 
(thin curves) as a function of the cusp indicator
$\chi$ along sequences with the adiabatic index $\gamma=2.5$.
The left (right) panel is for synchronized (irrotational) binaries.
Solid, dotted, dashed, and long-dashed curves correspond to 
$M/R=0.14~{\rm vs}~0.14$, 0.16 vs 0.16, 0.18 vs 0.18, and 0.20 vs 0.20,
respectively.}
\end{figure}%

\begin{figure}[htb]
\vspace{7mm}
\centerline{\includegraphics[width=0.4\textwidth]{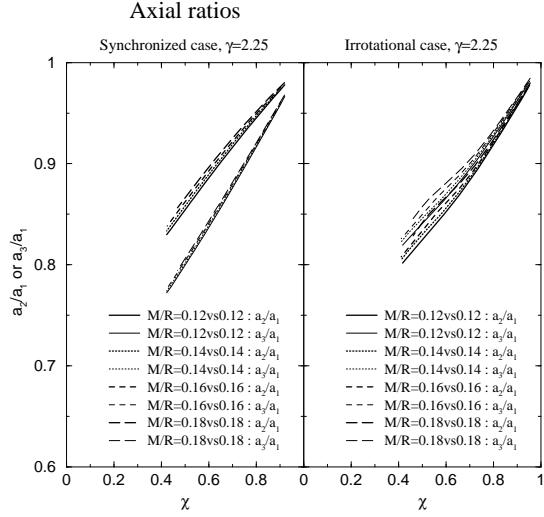}}
\caption[]{\label{fig:axis_eq_g225}
Same as Fig. \ref{fig:axis_eq_g25} but for $\gamma=2.25$.
Solid, dotted, dashed, and long-dashed curves correspond to 
$M/R=0.12~{\rm vs}~0.12$, 0.14 vs 0.14, 0.16 vs 0.16, and 0.18 vs 0.18,
respectively.
}
\end{figure}%

\begin{figure}[htb]
\vspace{7mm}
\centerline{\includegraphics[width=0.4\textwidth]{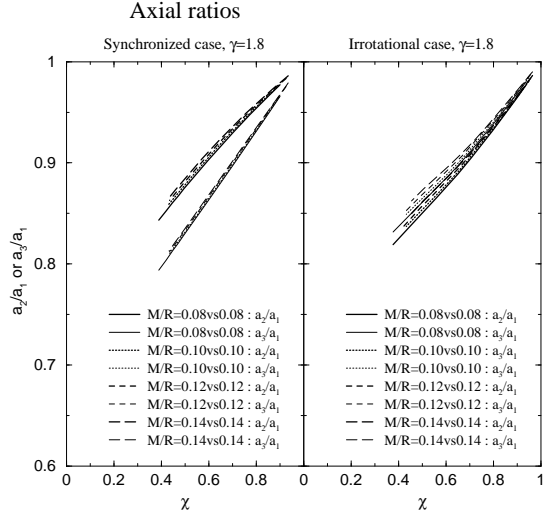}}
\caption[]{\label{fig:axis_eq_g18}
Same as Fig. \ref{fig:axis_eq_g25} but for $\gamma=1.8$.
Solid, dotted, dashed, and long-dashed curves correspond to
$M/R=0.08~{\rm vs}~0.08$, 0.10 vs 0.10, 0.12 vs 0.12, and 0.14 vs 0.14,
respectively.
}
\end{figure}%

\begin{figure}[htb]
\vspace{7mm}
\centerline{\includegraphics[width=0.4\textwidth]{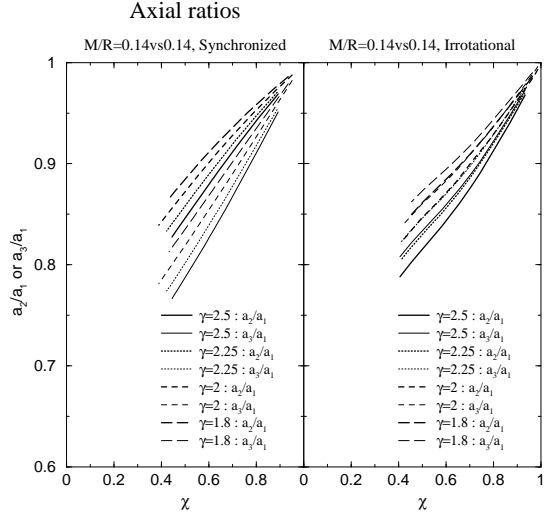}}
\caption[]{\label{fig:axis_gamma}
Axial ratios as a function of the cusp indicator $\chi$.
Left (right) is for synchronized (irrotational) binary systems
with compactness $M/R=0.14~{\rm vs}~0.14$.
Solid, dotted, dashed, and long-dashed curves correspond to
the adiabatic indices $\gamma=2.5$, 2.25, 2, and 1.8,
respectively.
Thick and thin curves are respectively 
for the axial ratios $a_2/a_1$ and $a_3/a_1$.
}
\end{figure}%

\subsection{Deformation of the stars}

The variation of the stellar shape along an evolutionary sequence
is presented in Figs. \ref{fig:axis_eq_g25}
-- \ref{fig:axis_eq_g18}.
In these figures, we depict the (coordinate) axial ratio  
$a_2/a_1$ and $a_3/a_1$ as a function of the cusp indicator $\chi$.
This allows one to speculate about
the values of the axial ratios at the end point of the sequence,
because the curves are close to straight lines.
Since $\chi=0$ corresponds to the end point of the sequence
(mass-shedding limit),
we can evaluate the deformation of the stars at the end of the sequence
by extrapolating the curves to $\chi=0$. The axial ratios at 
the end of the sequences are thus predicted to be 
$a_2/a_1 \sim 0.7$ and $a_3/a_1 \sim 0.6$ for synchronized binary
systems with $\gamma=2.5$, and $a_2/a_1 \sim 0.67 - 0.72$ and
$a_3/a_1 \sim 0.7 - 0.75$ for irrotational ones with $\gamma=2.5$.
It is also possible to perceive that for the same value of $\chi$
the stars with larger compactness are closer to the spherical figure
than those of smaller compactness.

We also present the change of the axial ratios along the sequence
in Fig. \ref{fig:axis_gamma} by fixing the compactness and varying
the adiabatic index.
It is clearly seen from this figure that that the deformation of the star
is larger for larger adiabatic index.

\subsection{Contours of baryon density}

\begin{figure}[htb]
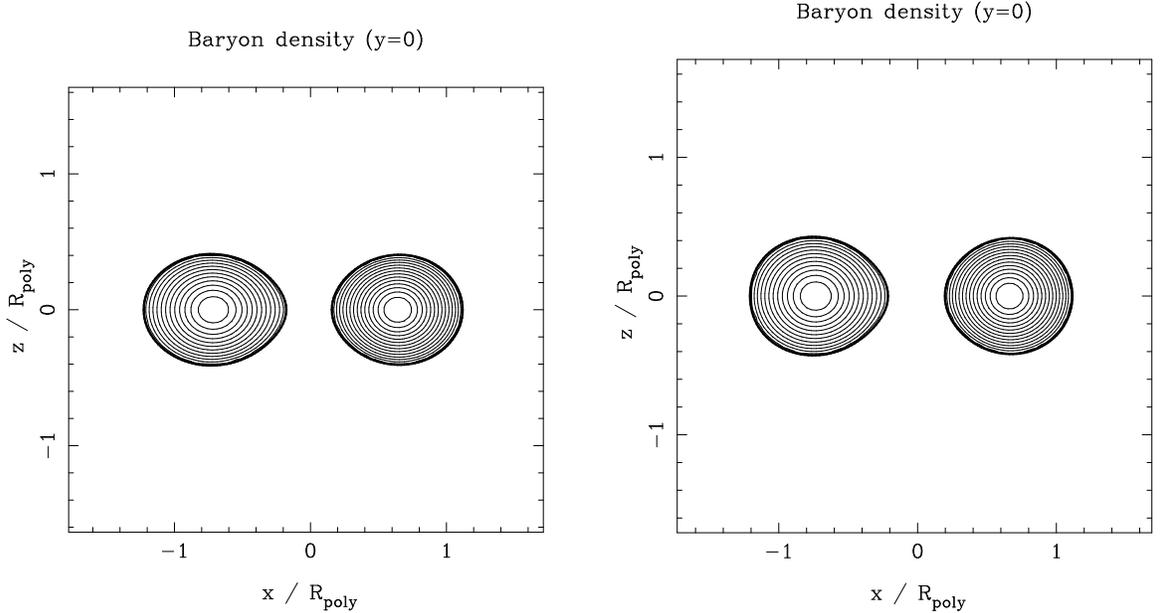

\centerline{\includegraphics[width=0.4\textwidth]{b_16vs18_cg25xz.eps}
  \hspace{20pt}
  \includegraphics[width=0.4\textwidth]{b_16vs18_ig25xz.eps}}
\caption[]{\label{fig:baryon_g25}
Isocontours of the baryon density for binary systems with
the adiabatic index $\gamma=2.5$ and the compactness $M/R=0.16~{\rm vs}~0.18$.
The left panel is for the synchronized case with
the orbital separation $d_G/R_{\rm poly} =1.356$,
while the right panel is for the irrotational one
with $d_G/R_{\rm poly} =1.393$.
The orbital axis is located on $x=0$ in each panel.
The thick solid lines denote the surface of the stars.
}
\end{figure}

\begin{figure}[htb]
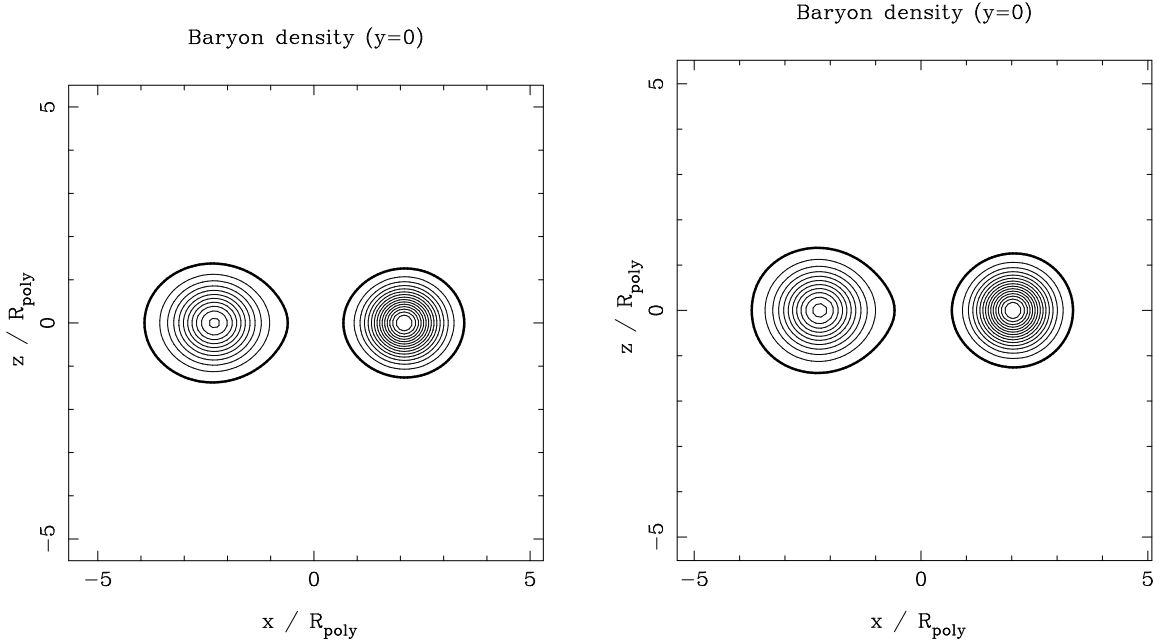

\centerline{\includegraphics[width=0.4\textwidth]{b_10vs12_cg18xz.eps}
  \hspace{20pt}
  \includegraphics[width=0.4\textwidth]{b_10vs12_ig18xz.eps}}
\caption[]{\label{fig:baryon_g18}
Isocontours of the baryon density for binary systems
with the adiabatic index $\gamma=1.8$ and the compactness
$M/R=0.10~{\rm vs}~0.12$.
The left panel is for the synchronized case with
the orbital separation $d_G/R_{\rm poly} =4.391$,
while the right panel is for the irrotational one
with $d_G/R_{\rm poly} =4.257$.
The orbital axis is located on $x=0$ in each panel.
}
\end{figure}

Isocontours of the baryon density of various binary systems are
shown in Figs. \ref{fig:baryon_g25} -- \ref{fig:baryon_g18}.
There are two panels for different-mass binary systems in each figure.
The left panel shows the result for the synchronized case
while the right one is that for the irrotational case.
In each panel, the left-hand side star is the less massive star.
It is clearly seen that the lighter star is tidally deformed
and elongated while the more massive star 
does not deviate very much from the spherical shape.


\begin{figure}[htb]
\centerline{\includegraphics[width=0.4\textwidth]{b_14vs16_cg225xz.eps}
  \hspace{20pt}
  \includegraphics[width=0.4\textwidth]{b_14vs16_ig225xz.eps}}
\caption[]{\label{fig:baryon_g225}
Isocontours of the baryon density for binary systems
with the adiabatic index $\gamma=2.25$ and the compactness
$M/R=0.14~{\rm vs}~0.16$.
The left panel is for the synchronized case with
the orbital separation $d_G/R_{\rm poly} =1.765$,
while the right panel is for the irrotational one
with $d_G/R_{\rm poly} =1.775$.
The orbital axis is located on $x=0$ in each panel.
}
\end{figure}%


\subsection{Tables of results}

Finally, we summarize our numerical results about constant baryon number
sequences in Tables \ref{table1a} -- \ref{table3d}.
In each table, we first give the orbital separation
defined as the distance $d_G$
between the two ``center of mass'' (see Eq.~(128) of Paper~I for a precise
definition), in  polytropic units:
\beq
  \bar{d}_G := {d_G \over R_{\rm poly}},
\eeq
where $R_{\rm poly}$ is the length constructed from the 
the polytropic constant $\kappa$ and the adiabatic index $\gamma$:
\beq
  R_{\rm poly} := \kappa^{1/2(\gamma-1)} . 
\eeq
We also list in Tables \ref{table1a} -- \ref{table3d}
the relative separation $\tilde d$ defined by Eq.~(\ref{e:def_tilde_d})
(let us recall that ${\tilde d}=1$ indicates the contact between the 
two stars).
Next we give the dimensionless orbital angular velocity:
\beq
  \bar{\Omega} := \Omega \kappa^{1/2(\gamma-1)},
\eeq
the dimensionless ADM mass:
\beq
  \bar{M} := M \kappa^{-1/2(\gamma-1)},
\eeq
the dimensionless baryon mass:
\beq
  \bar{M}_{\rm B} := M_{\rm B} \kappa^{-1/2(\gamma-1)},
\eeq
and the dimensionless total angular momentum:
\beq
  \bar{J} := J \kappa^{-1/(\gamma-1)}.
\eeq
The absolute values of the relative errors in the virial theorem,
$|VE(M)|$, $|VE(FUS)|$, and $|VE(GB)|$, are defined by Eqs.
(\ref{eq:virial_mass}) -- (\ref{eq:virial_gb}).
$\d e_c$ is the relative change in central energy density defined 
by Eq.~(\ref{e:rel_e_c}). 
In the tables, primes denote values for the more massive star,
and the symbol $\dagger$ denotes the turning point in the ADM mass
(and the total angular momentum) along the sequence.

\section{Discussion} \label{s:discussion}

\begin{figure}[htb]
\vspace{7mm}
\centerline{\includegraphics[width=0.4\textwidth]{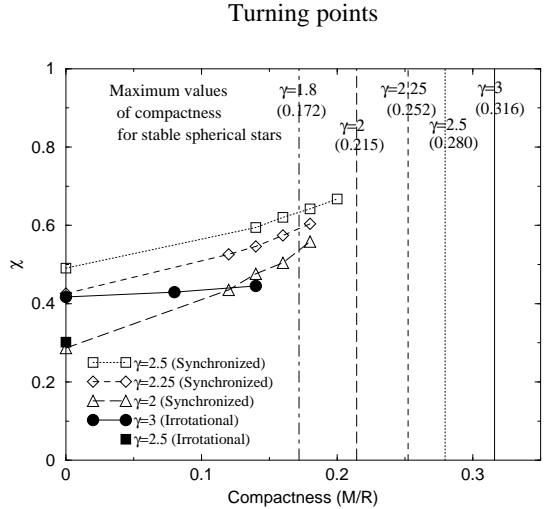}}
\caption[]{\label{fig:turning}
Positions of the turning points of the ADM mass (and total angular
momentum) in the sequence as a function of the compactness parameter.
Solid curve with filled circle is for the irrotational binary system
with $\gamma=3$.
Dotted with open square, dashed with open diamond, and long-dashed
with open triangle denote the results of synchronized binary systems
with $\gamma=2.5$, 2.25, and 2, respectively.
Vertical solid, dotted, dashed, and long-dashed lines are
maximum values of compactness for stable spherical stars (at $\chi=1$)
for $\gamma=3$, 2.5, 2.25, 2, and 1.8, respectively.
}
\end{figure}

As shown by Friedman et al. \cite{FriedUS02}, the minimum (turning point)
in the ADM mass in a constant baryon number sequence locates 
the appearance of a secular (resp. dynamical) instability for 
synchronized (resp. irrotational) binaries, thus defining
the {\em innermost stable circular orbit (ISCO)}. 
The frequency at the ISCO is a potentially observable parameter
by the gravitational wave detectors, and thus a very interesting
quantity.  
The position of the ISCO for identical-mass binary systems
is depicted as a function of the compactness parameter
in Fig. \ref{fig:turning}.
In this figure, the Newtonian results are shown as $M/R=0$.
The corresponding values of $\chi$ for the irrotational binaries
are slightly different from those in Paper II.
This is because we have improved our numerical code
to decrease the aliasing error
which appears in the spectral method \cite{CanutHQZ88}.
 
Several things can be found and predicted from Fig.~\ref{fig:turning}.
Firstly, the turning points are found for synchronized binary systems
for $\gamma \ge 2$ and for irrotational ones for $\gamma \ge 2.5$.
This last finding agrees with the results of Uryu et al. \cite{UryuSE00}.
The turning point might exist for adiabatic index slightly lower than 2
for synchronized binary systems and 2.5 for irrotational ones.
However, since it is very difficult for our numerical code to calculate
quasiequilibrium figures of $\chi < 0.4$ for mildly relativistic cases
and $\chi < 0.5$ for highly relativistic ones,
we cannot determine the exact value of the minimum adiabatic index
which has the turning point.
Secondly, the quantity $\chi$ at the turning point increases proportionally
to the compactness of the stars.
This implies that the relative orbital separation at the turning point
increases with the increase of the compactness.
We should comment here that
the tendency for the irrotational binary system with $\gamma=3$
was formerly given by Uryu et al. \cite{UryuSE00}.
Thirdly, the inclination of lines for synchronized binary systems
is larger than that for irrotational ones.
Fourthly, we may assert that a turning point is found
in Newtonian calculations for some value of the adiabatic index $\gamma$,
it should exist in the general relativistic computations
(within the IWM approximation) for the same value of $\gamma$,
because the lines in Fig.~\ref{fig:turning} have a positive slope. 
It is worth to point out that even if the turning point is not found
in the Newtonian sequences,
the possibility to find it in the general relativistic computations
does not vanish.
In Fig. \ref{fig:turning}, some vertical lines which
denote the maximum values of compactness for stable spherical stars
are depicted.
They show the limit of the existence of the turning point for
each adiabatic index.
Then, for a given adiabatic index, the turning point could exist from a
compactness $M/R > 0$ to the maximum $(M/R)_{\rm max}$
taking into account the fact that the line of the turning point
has a positive inclination.

\section{Summary} \label{s:summary}

We have developed a numerical code for computing quasiequilibrium
sequences of binary neutron stars through the series of works
\cite{GourGTMB01,TanigGB01,TanigG02a,TanigG02b}.
In the present article, we have presented the results of
relativistic computations of both
synchronized and irrotational rotation states, 
for polytropic equations of state with several adiabatic indices.
We have considered not only binary systems composed of
identical-mass stars but also different-mass systems. 
We have investigated the behaviors of various physical 
quantities along constant
baryon number sequences (evolutionary sequences): 
the relative change in central energy density,
the ADM mass, the total angular momentum, the shape of the figures,
as well as the location of the end point and that 
of the turning point in the ADM mass and angular momentum.
 
As one of interesting results,
we propose, as a conjecture, that if the turning points defining 
the ISCO are found
for the Newtonian calculations for some adiabatic index $\gamma$,
they should exist in the general relativistic computations
for the same value of $\gamma$.
Of course, this conjecture is derived from the results of non-realistic
cases such as synchronized binary systems or irrotational ones
with $\gamma=3$.
Furthermore, since the inclination of lines in Fig. \ref{fig:turning}
is positive,
it is not possible to exactly predict whether or not a turning point
appears in the IWM approximation
when there is no turning point in the Newtonian calculation.
However, if the computations of irrotational binary systems
for $\chi<0.2$ could be obtained in the future
with sufficient accuracy in Newtonian gravity,
which is easier to be calculated than in the relativistic framework,
it would be possible to speculate
about the turning points in the IWM approximation without computations.

The numerical results presented in this article
are freely available from
http://www.lorene.obspm.fr/data/
This web page provides binary data files containing
the whole binary neutron stars configurations
as well as some codes to read them.

\acknowledgments

We would like to thank Koji Uryu for usefull discussion and providing
us with some unpublished data. 
KT acknowledges a Grant-in-Aid for Scientific Research (No. 14-06898)
of the Japanese Ministry of Education, Culture, Sports,
Science and Technology.


\newpage

\begin{table}
\caption{Comparison with the results of Uryu, Shibata and Eriguchi
\cite{UryuSE00} and Shibata and Uryu \cite{ShibaU01,Uryu03}.
}
 \begin{ruledtabular}
\begin{tabular}{lllllll}
  &$M/R$&$\bar{M}_{\rm B}$&$0.5 \times \bar{M}$&$M_{\rm B} \Omega$&
  $J/M_{\rm ADM}^2$&$d_G/M_{\rm ADM}$ \\ \hline
  \multicolumn{7}{c}{$\gamma=2.5$} \\
  this work&0.14&0.08742&0.07974&0.01836&0.9563&8.815 \\
  Uryu et al.    &0.14&0.0874 &0.0797 &0.0189 &0.981 &8.56 \\
  \multicolumn{7}{c}{$\gamma=2.25$} \\
  this work&0.12&0.09316&0.08639&0.01414&1.004&10.42 \\
  Uryu et al.    &0.12&0.0932 &0.0864 &0.0145 &1.03 &10.3 \\
  this work&0.14&0.1080&0.09886&0.01825&0.9521&8.812 \\
  Uryu et al.    &0.14&0.108 &0.0988 &0.0189 &0.975 &8.52 \\
  \multicolumn{7}{c}{$\gamma=2$} \\
  this work&0.12&0.1299&0.1211&0.01423&0.9972&10.42 \\
  Uryu et al.    &0.12&0.130 &0.121 &0.0144 &1.02  &10.3 \\
  this work&0.14&0.1461&0.1347&0.01809&0.9474&8.805 \\
  Uryu et al.    &0.14&0.146 &0.135 &0.0187 &0.971 &8.52 \\
  Shibata \& Uryu&0.14&0.14614&0.1347& 0.01885 &0.9419& 8.554  \\
  \multicolumn{7}{c}{$\gamma=1.8$} \\
  this work&0.10&0.1725&0.1635&0.01038&1.065&12.88 \\
  Uryu et al.    &0.10&0.173 &0.164 &0.0106 &1.08 &12.6 \\
  this work&0.12&0.1920&0.1803&0.01394&0.9947&10.49 \\
  Uryu et al.    &0.12&0.192 &0.180 &0.0143 &1.02  &10.2 \\
  this work&0.14&0.2065&0.1924&0.01767&0.9456&8.871 \\
  Uryu et al.    &0.14&0.207 &0.192 &0.0185 &0.967 &8.50 \\
  \end{tabular}
 \end{ruledtabular}
 \label{table0}
\end{table}%


\begin{table}
\caption{Orbital angular velocity, ADM mass, total angular momentum,
virial errors, axial ratios, relative change in central energy density,
and the cusp indicator $\chi$ along constant baryon number sequences
of identical mass binaries in synchronized motion, for various compactness.
The symbol $\dagger$ denotes the turning point of the sequences.
}
 \begin{ruledtabular}
  \begin{tabular}{rccccccccccc}
  \multicolumn{12}{c}{Identical mass stars, Synchronized case, $\gamma=2.5$}
  \\
  $\bar{d}_G$&$\tilde{d}$&$\bar{\Omega}$&$\bar{M}$&$\bar{J}$&
  $|VE(M)|$&$|VE(GB)|$&$|VE(FUS)|$&
  $a_2/a_1$&$a_3/a_1$&$\d e_{\rm c}$&$\chi$ \\ \hline
   \multicolumn{12}{c}
  {$M/R=0.14~{\rm vs}~0.14,~~\bar{M}_{\rm B}=0.08742~{\rm vs.}~0.08742$} \\
  2.186&2.238&0.1126&0.1601&2.964(-2)&1.174(-5)&2.242(-5)&2.123(-5)&
       0.9689&0.9512&-1.397(-2)&0.8927 \\
  1.822&1.829&0.1452&0.1599&2.857(-2)&8.314(-6)&1.627(-5)&1.529(-5)&
       0.9443&0.9134&-2.447(-2)&0.8072 \\
  1.640&1.607&0.1681&0.1599&2.819(-2)&6.026(-6)&1.204(-5)&1.130(-5)&
       0.9187&0.8771&-3.453(-2)&0.7232 \\
  1.531&1.461&0.1851&0.1599&2.806(-2)&3.095(-5)&5.471(-5)&5.052(-5)&
       0.8924&0.8426&-4.397(-2)&0.6408 \\
$\dagger$
  1.487&1.398&0.1928&0.1599&2.805(-2)&2.776(-5)&4.804(-5)&4.397(-5)&
       0.8774&0.8240&-4.897(-2)&0.5948 \\
\end{tabular}
 \end{ruledtabular}
 \label{table1a}
\end{table}%

\begin{table}
\caption{Same as Table \ref{table1a} but for different mass binary systems.
The prime denotes the values for the more massive star.
}
 \begin{ruledtabular}
  \begin{tabular}{rccccccccccc}
  \multicolumn{12}{c}{Different mass stars, Synchronized case, $\gamma=2.5$}
  \\
  $\bar{d}_G$&$\tilde{d}$&$\bar{\Omega}$&$\bar{M}$&$\bar{J}$&
  $|VE(M)|$&$|VE(GB)|$&$|VE(FUS)|$&
  $a_2/a_1$&$a_3/a_1$&$\d e_{\rm c}$&$\chi$ \\
   & & & & & & & &$a_2'/a_1'$&$a_3'/a_1'$&$\d e_{\rm c}'$&$\chi'$  \\ \hline
  \multicolumn{12}{c}
  {$M/R=0.14~{\rm vs}~0.16,~~\bar{M}_{\rm B}=0.08742~{\rm vs.}~0.1005$} \\
  2.004&2.080&0.1305&0.1708&3.224(-2)&1.331(-5)&2.585(-5)&2.443(-5)&
          0.9535&0.9300&-1.939(-2)&0.8452 \\
  &&&&&&&&0.9691&0.9480&-1.740(-2)&0.8833 \\
  1.822&1.870&0.1489&0.1707&3.170(-2)&1.120(-5)&2.191(-5)&2.066(-5)&
          0.9365&0.9051&-2.617(-2)&0.7888 \\
  &&&&&&&&0.9581&0.9300&-2.318(-2)&0.8419 \\
\end{tabular}
 \end{ruledtabular}
 \label{table1b}
\end{table}%

\begin{table}
\caption{Same as Table \ref{table1a} but for irrotational binaries.
}
 \begin{ruledtabular}
  \begin{tabular}{rccccccccccc}
  \multicolumn{12}{c}{Identical mass stars, Irrotational case, $\gamma=2.5$}
  \\
  $\bar{d}_G$&$\tilde{d}$&$\bar{\Omega}$&$\bar{M}$&$\bar{J}$&
  $|VE(M)|$&$|VE(GB)|$&$|VE(FUS)|$&
  $a_2/a_1$&$a_3/a_1$&$\d e_{\rm c}$&$\chi$ \\ \hline
   \multicolumn{12}{c}
  {$M/R=0.14~{\rm vs}~0.14,~~\bar{M}_{\rm B}=0.08742~{\rm vs.}~0.08742$} \\
  2.185&2.266&0.1129&0.1600&2.765(-2)&1.044(-5)&1.994(-5)&1.865(-5)&
       0.9678&0.9729&-4.680(-4)&0.9344 \\
  1.821&1.865&0.1459&0.1598&2.599(-2)&6.162(-6)&1.208(-5)&1.095(-5)&
       0.9403&0.9474&-1.371(-3)&0.8693 \\
  1.565&1.552&0.1803&0.1596&2.487(-2)&1.366(-5)&2.090(-5)&1.565(-6)&
       0.8903&0.9018&-3.843(-3)&0.7422 \\
  1.491&1.446&0.1930&0.1595&2.458(-2)&2.819(-5)&5.056(-5)&4.520(-5)&
       0.8608&0.8751&-5.587(-3)&0.6573 \\
  1.454&1.383&0.1999&0.1595&2.445(-2)&2.273(-5)&4.210(-5)&3.612(-5)&
       0.8384&0.8548&-6.944(-3)&0.5839 \\
\end{tabular}
 \end{ruledtabular}
 \label{table1c}
\end{table}%

\begin{table}
\caption{Same as Table \ref{table1c} but for different mass binary systems.
The prime denotes the values for the more massive star.
}
 \begin{ruledtabular}
  \begin{tabular}{rccccccccccc}
  \multicolumn{12}{c}{Different mass stars, Irrotational case, $\gamma=2.5$}
  \\
  $\bar{d}_G$&$\tilde{d}$&$\bar{\Omega}$&$\bar{M}$&$\bar{J}$&
  $|VE(M)|$&$|VE(GB)|$&$|VE(FUS)|$&
  $a_2/a_1$&$a_3/a_1$&$\d e_{\rm c}$&$\chi$ \\
   & & & & & & & &$a_2'/a_1'$&$a_3'/a_1'$&$\d e_{\rm c}'$&$\chi'$  \\ \hline
  \multicolumn{12}{c}
  {$M/R=0.14~{\rm vs}~0.16,~~\bar{M}_{\rm B}=0.08742~{\rm vs.}~0.1005$} \\
  2.003&2.112&0.1309&0.1707&2.978(-2)&1.129(-5)&2.193(-5)&2.036(-5)&
          0.9509&0.9581&-1.012(-3)&0.8976 \\
  &&&&&&&&0.9677&0.9725&-5.307(-4)&0.9307 \\
  1.821&1.906&0.1496&0.1705&2.888(-2)&8.442(-6)&1.665(-5)&1.521(-5)&
          0.9316&0.9402&-1.799(-3)&0.8516 \\
  &&&&&&&&0.9553&0.9609&-9.288(-4)&0.9004 \\
\end{tabular}
 \end{ruledtabular}
 \label{table1d}
\end{table}%


\begin{table}
\caption{Same as Table \ref{table1a} but for $\gamma=2.25$.
}
 \begin{ruledtabular}
  \begin{tabular}{rccccccccccc}
  \multicolumn{12}{c}{Identical mass stars, Synchronized case, $\gamma=2.25$}
  \\
  $\bar{d}_G$&$\tilde{d}$&$\bar{\Omega}$&$\bar{M}$&$\bar{J}$&
  $|VE(M)|$&$|VE(GB)|$&$|VE(FUS)|$&
  $a_2/a_1$&$a_3/a_1$&$\d e_{\rm c}$&$\chi$ \\ \hline
  \multicolumn{12}{c}
  {$M/R=0.12~{\rm vs}~0.12,~~\bar{M}_{\rm B}=0.09316~{\rm vs.}~0.09316$} \\
  2.925&2.304&0.07708&0.1734&3.724(-2)&8.099(-6)&1.407(-5)&1.317(-5)&
       0.9726&0.9578&-1.402(-2)&0.9002 \\
  2.701&2.113&0.08635&0.1733&3.646(-2)&7.249(-6)&1.259(-5)&1.174(-5)&
       0.9649&0.9457&-1.787(-2)&0.8712 \\
  2.251&1.714&0.1117&0.1732&3.510(-2)&4.990(-6)&8.736(-6)&8.072(-6)&
       0.9354&0.9019&-3.171(-2)&0.7650 \\
  2.026&1.492&0.1296&0.1731&3.462(-2)&2.062(-5)&3.400(-5)&3.093(-5)&
       0.9027&0.8580&-4.536(-2)&0.6549 \\
  1.936&1.392&0.1383&0.1731&3.452(-2)&1.853(-5)&2.870(-5)&2.572(-5)&
       0.8808&0.8307&-5.355(-2)&0.5840 \\
\end{tabular}
 \end{ruledtabular}
 \label{table2a}
\end{table}%

\begin{table}
\caption{Same as Table \ref{table2a} but for different mass binary systems.
The prime denotes the values for the more massive star.
}
 \begin{ruledtabular}
  \begin{tabular}{rccccccccccc}
  \multicolumn{12}{c}{Different mass stars, Synchronized case, $\gamma=2.25$}
  \\
  $\bar{d}_G$&$\tilde{d}$&$\bar{\Omega}$&$\bar{M}$&$\bar{J}$&
  $|VE(M)|$&$|VE(GB)|$&$|VE(FUS)|$&
  $a_2/a_1$&$a_3/a_1$&$\d e_{\rm c}$&$\chi$ \\
   & & & & & & & &$a_2'/a_1'$&$a_3'/a_1'$&$\d e_{\rm c}'$&$\chi'$  \\ \hline
  \multicolumn{12}{c}
  {$M/R=0.12~{\rm vs}~0.14,~~\bar{M}_{\rm B}=0.09316~{\rm vs.}~0.1080$} \\
  2.701&2.169&0.08889&0.1859&4.063(-2)&9.821(-6)&1.738(-5)&1.630(-5)&
          0.9595&0.9400&-1.916(-2)&0.8577 \\
  &&&&&&&&0.9744&0.9576&-1.665(-2)&0.8980 \\
  2.476&1.970&0.1005&0.1858&3.983(-2)&8.511(-6)&1.508(-5)&1.407(-5)&
          0.9464&0.9208&-2.513(-2)&0.8113 \\
  &&&&&&&&0.9663&0.9443&-2.161(-2)&0.8655 \\
\end{tabular}
 \end{ruledtabular}
 \label{table2b}
\end{table}%

\begin{table}
\caption{Same as Table \ref{table2a} but for irrotational binaries.
}
 \begin{ruledtabular}
  \begin{tabular}{rccccccccccc}
  \multicolumn{12}{c}{Identical mass stars, Irrotational case, $\gamma=2.25$}
  \\
  $\bar{d}_G$&$\tilde{d}$&$\bar{\Omega}$&$\bar{M}$&$\bar{J}$&
  $|VE(M)|$&$|VE(GB)|$&$|VE(FUS)|$&
  $a_2/a_1$&$a_3/a_1$&$\d e_{\rm c}$&$\chi$ \\ \hline
  \multicolumn{12}{c}
  {$M/R=0.12~{\rm vs}~0.12,~~\bar{M}_{\rm B}=0.09316~{\rm vs.}~0.09316$} \\
  2.924&2.331&0.07722&0.1733&3.503(-2)&7.443(-6)&1.278(-5)&1.184(-5)&
       0.9720&0.9764&-3.975(-4)&0.9392 \\
  2.699&2.145&0.08654&0.1732&3.399(-2)&6.429(-6)&1.096(-6)&1.004(-5)&
       0.9639&0.9689&-6.208(-4)&0.9189 \\
  2.249&1.756&0.1121&0.1730&3.186(-2)&3.645(-6)&5.956(-6)&5.181(-6)&
       0.9318&0.9390&-1.880(-3)&0.8364 \\
  2.023&1.540&0.1302&0.1729&3.081(-2)&1.034(-5)&1.390(-5)&1.027(-5)&
       0.8949&0.9051&-3.891(-3)&0.7350 \\
  1.933&1.440&0.1390&0.1728&3.041(-2)&1.791(-5)&2.913(-5)&2.545(-5)&
       0.8689&0.8813&-5.503(-3)&0.6559 \\
\end{tabular}
 \end{ruledtabular}
 \label{table2c}
\end{table}%

\begin{table}
\caption{Same as Table \ref{table2c} but for different mass binary systems.
The prime denotes the values for the more massive star.
}
 \begin{ruledtabular}
  \begin{tabular}{rccccccccccc}
  \multicolumn{12}{c}{Different mass stars, Irrotational case, $\gamma=2.25$}
  \\
  $\bar{d}_G$&$\tilde{d}$&$\bar{\Omega}$&$\bar{M}$&$\bar{J}$&
  $|VE(M)|$&$|VE(GB)|$&$|VE(FUS)|$&
  $a_2/a_1$&$a_3/a_1$&$\d e_{\rm c}$&$\chi$ \\
   & & & & & & & &$a_2'/a_1'$&$a_3'/a_1'$&$\d e_{\rm c}'$&$\chi'$  \\ \hline
  \multicolumn{12}{c}
  {$M/R=0.12~{\rm vs}~0.14,~~\bar{M}_{\rm B}=0.09316~{\rm vs.}~0.1080$} \\
  2.699&2.200&0.08910&0.1858&3.794(-2)&8.995(-6)&1.576(-5)&1.458(-5)&
          0.9582&0.9643&-8.350(-4)&0.9072 \\
  &&&&&&&&0.9737&0.9778&-4.042(-4)&0.9404 \\
  2.474&2.006&0.1008&0.1857&3.678(-2)&7.210(-6)&1.246(-5)&1.134(-5)&
          0.9440&0.9512&-1.386(-3)&0.8716 \\
  &&&&&&&&0.9651&0.9697&-6.597(-4)&0.9182 \\
\end{tabular}
 \end{ruledtabular}
 \label{table2d}
\end{table}%


\begin{table}
\caption{Same as Table \ref{table1a} but for $\gamma=1.8$.
}
 \begin{ruledtabular}
  \begin{tabular}{rccccccccccc}
  \multicolumn{12}{c}{Identical mass stars, Synchronized case, $\gamma=1.8$}
  \\
  $\bar{d}_G$&$\tilde{d}$&$\bar{\Omega}$&$\bar{M}$&$\bar{J}$&
  $|VE(M)|$&$|VE(GB)|$&$|VE(FUS)|$&
  $a_2/a_1$&$a_3/a_1$&$\d e_{\rm c}$&$\chi$ \\ \hline
  \multicolumn{12}{c}
  {$M/R=0.08~{\rm vs}~0.08,~~\bar{M}_{\rm B}=0.1485~{\rm vs.}~0.1485$} \\
  7.747&2.340&0.02361&0.2852&0.1200&3.257(-6)&4.951(-6)&4.521(-6)&
       0.9771&0.9657&-1.777(-2)&0.9005 \\
  6.887&2.058&0.02800&0.2850&0.1155&2.891(-6)&4.363(-6)&3.941(-6)&
       0.9668&0.9499&-2.546(-2)&0.8546 \\
  6.026&1.763&0.03396&0.2849&0.1112&2.427(-6)&3.637(-6)&3.258(-6)&
       0.9475&0.9218&-3.873(-2)&0.7729 \\
  5.596&1.605&0.03779&0.2848&0.1092&2.044(-6)&3.034(-6)&2.689(-6)&
       0.9305&0.8985&-4.931(-2)&0.7047 \\
  5.165&1.434&0.04243&0.2848&0.1074&3.124(-6)&5.190(-6)&4.906(-6)&
       0.9026&0.8625&-6.482(-2)&0.5977 \\
\end{tabular}
 \end{ruledtabular}
 \label{table3a}
\end{table}%

\begin{table}
\caption{Same as Table \ref{table3a} but for different mass binary systems.
The prime denotes the values for the more massive star.
}
 \begin{ruledtabular}
  \begin{tabular}{rccccccccccc}
  \multicolumn{12}{c}{Different mass stars, Synchronized case, $\gamma=1.8$}
  \\
  $\bar{d}_G$&$\tilde{d}$&$\bar{\Omega}$&$\bar{M}$&$\bar{J}$&
  $|VE(M)|$&$|VE(GB)|$&$|VE(FUS)|$&
  $a_2/a_1$&$a_3/a_1$&$\d e_{\rm c}$&$\chi$ \\
   & & & & & & & &$a_2'/a_1'$&$a_3'/a_1'$&$\d e_{\rm c}'$&$\chi'$  \\ \hline
  \multicolumn{12}{c}
  {$M/R=0.08~{\rm vs}~0.10,~~\bar{M}_{\rm B}=0.1485~{\rm vs.}~0.1725$} \\
  7.747&2.473&0.02440&0.3067&0.1335&5.001(-6)&7.956(-6)&7.382(-6)&
          0.9734&0.9617&-1.912(-2)&0.8890 \\
  &&&&&&&&0.9860&0.9781&-1.478(-2)&0.9354 \\
  6.887&2.178&0.02892&0.3066&0.1285&4.532(-6)&7.170(-6)&6.587(-6)&
          0.9614&0.9440&-2.745(-2)&0.8377 \\
  &&&&&&&&0.9801&0.9683&-2.097(-2)&0.9064 \\
\end{tabular}
 \end{ruledtabular}
 \label{table3b}
\end{table}%

\begin{table}
\caption{Same as Table \ref{table3a} but for irrotational binaries.
}
 \begin{ruledtabular}
  \begin{tabular}{rccccccccccc}
  \multicolumn{12}{c}{Identical mass stars, Irrotational case, $\gamma=1.8$}
  \\
  $\bar{d}_G$&$\tilde{d}$&$\bar{\Omega}$&$\bar{M}$&$\bar{J}$&
  $|VE(M)|$&$|VE(GB)|$&$|VE(FUS)|$&
  $a_2/a_1$&$a_3/a_1$&$\d e_{\rm c}$&$\chi$ \\ \hline
  \multicolumn{12}{c}
  {$M/R=0.08~{\rm vs}~0.08,~~\bar{M}_{\rm B}=0.1485~{\rm vs.}~0.1485$} \\
  7.746&2.369&0.02363&0.2851&0.1147&3.139(-6)&4.714(-6)&4.269(-6)&
       0.9772&0.9804&-3.530(-4)&0.9387 \\
  6.024&1.812&0.03401&0.2848&0.1033&2.178(-6)&3.112(-6)&2.689(-6)&
       0.9480&0.9527&-1.476(-3)&0.8488 \\
  5.593&1.663&0.03785&0.2847&0.1004&1.839(-6)&2.535(-6)&2.128(-6)&
       0.9317&0.9373&-2.346(-3)&0.7966 \\
  5.162&1.504&0.04251&0.2845&0.09736&7.196(-6)&9.673(-6)&8.089(-6)&
       0.9050&0.9122&-4.010(-3)&0.7077 \\
  4.989&1.433&0.04465&0.2845&0.09616&7.030(-6)&9.188(-6)&7.585(-6)&
       0.8887&0.8968&-5.091(-3)&0.6494 \\
\end{tabular}
 \end{ruledtabular}
 \label{table3c}
\end{table}%

\begin{table}
\caption{Same as Table \ref{table3c} but for different mass binary systems.
The prime denotes the values of the more massive star.
}
 \begin{ruledtabular}
  \begin{tabular}{rccccccccccc}
  \multicolumn{12}{c}{Different mass stars, Irrotational case, $\gamma=1.8$}
  \\
  $\bar{d}_G$&$\tilde{d}$&$\bar{\Omega}$&$\bar{M}$&$\bar{J}$&
  $|VE(M)|$&$|VE(GB)|$&$|VE(FUS)|$&
  $a_2/a_1$&$a_3/a_1$&$\d e_{\rm c}$&$\chi$ \\
   & & & & & & & &$a_2'/a_1'$&$a_3'/a_1'$&$\d e_{\rm c}'$&$\chi'$  \\ \hline
  \multicolumn{12}{c}
  {$M/R=0.08~{\rm vs}~0.10,~~\bar{M}_{\rm B}=0.1485~{\rm vs.}~0.1725$} \\
  6.885&2.215&0.02895&0.3065&0.1219&4.305(-6)&6.716(-6)&6.103(-6)&
          0.9616&0.9662&-9.151(-4)&0.8936 \\
  &&&&&&&&0.9802&0.9832&-3.373(-4)&0.9453 \\
  6.024&1.920&0.03510&0.3063&0.1155&3.552(-6)&5.429(-6)&4.824(-6)&
          0.9395&0.9453&-1.994(-3)&0.8253 \\
  &&&&&&&&0.9697&0.9732&-6.930(-4)&0.9125 \\
\end{tabular}
 \end{ruledtabular}
 \label{table3d}
\end{table}%

\end{document}